\documentclass[%
reprint,
superscriptaddress,
amsmath,
amssymb,aps,
prc
]{revtex4-2}

\usepackage{graphicx}
\usepackage{dcolumn}
\usepackage{bm}
\usepackage{enumitem}
\usepackage{xcolor}

\begin{document}

\newcommand{\hiro}[1]{\textcolor{green}{#1}}
\newcommand{\ig}[1]{\textcolor{red}{#1}}
\newcommand{\sg}[1]{\textcolor{red}{#1}}

\preprint{APS/123-QED}

\title{Giant Dipole Resonance photofission and photoneutron reactions in $^{238}$U and $^{232}$Th}

\author{D.~Filipescu}\email{dan.filipescu@nipne.ro}
\affiliation{National Institute for Physics and Nuclear Engineering,
Horia Hulubei (IFIN-HH), 30 Reactorului, 077125 Bucharest-Magurele, Romania}

\author{I.~Gheorghe} 
\affiliation{National Institute for Physics and Nuclear Engineering,
Horia Hulubei (IFIN-HH), 30 Reactorului, 077125 Bucharest-Magurele, Romania}

\author{S.~Goriely} 
\affiliation{Institut d'Astronomie et d'Astrophysique, Universit\'e Libre de Bruxelles, Campus de la Plaine CP 226, 1050 Brussels, Belgium}

\author{A.~Tudora} 
\affiliation{University of Bucharest, Faculty of Physics, str. Atomistilor 405, Bucharest-Magurele, Jud. Ilfov, 077125, Romania}

\author{K.~Nishio} 
\affiliation{Advanced Science Research Center, Japan Atomic Energy Agency, Tokai, Ibaraki 319-1195 Japan}

\author{T.~Ohtsuki} 
\affiliation{Institute for Integrated Radiation and Nuclear Science, Kyoto University,
2-1010 Asashiro-nishi, Kumatori, Sennan, Osaka 590-0494, Japan}

\author{H.~Wang} 
\affiliation{Shanghai Advanced Research Institute, Chinese Academy of Sciences, No.99 Haike Road, Zhangjiang Hi-Tech Park, 201210 Pudong Shanghai, China}

\author{G.~Fan} 
\affiliation{Shanghai Advanced Research Institute, Chinese Academy of Sciences, No.99 Haike Road, Zhangjiang Hi-Tech Park, 201210 Pudong Shanghai, China}

\author{K.~Stopani} 
\affiliation{Lomonosov Moscow State University, Skobeltsyn Institute of Nuclear Physics, 119991 Moscow, Russia}

\author{F.~Suzaki} 
\affiliation{Advanced Science Research Center, Japan Atomic Energy Agency, Tokai, Ibaraki 319-1195 Japan}

\author{K.~Hirose} 
\affiliation{Advanced Science Research Center, Japan Atomic Energy Agency, Tokai, Ibaraki 319-1195 Japan}

\author{M.~Inagaki} 
\affiliation{Institute for Integrated Radiation and Nuclear Science, Kyoto University,
2-1010 Asashiro-nishi, Kumatori, Sennan, Osaka 590-0494, Japan}


\author{Y.-W.~Lui} 
\affiliation{Cyclotron Institute, Texas A\&M University, College Station, Texas 77843, USA}

\author{T.~Ari-izumi} 
\affiliation{Konan University, Department of Physics, 8-9-1 Okamoto, Higashinada, Kobe 658-8501, Japan}

\author{S.~Miyamoto} 
\affiliation{Laboratory of Advanced Science and Technology for Industry, University of Hyogo, 3-1-2 Kouto, Kamigori, Ako-gun, Hyogo 678-1205, Japan}

\author{T.~Otsuka} 
\affiliation{Department of Physics, The University of Tokyo, 7-3-1 Hongo, Bunkyo, Tokyo 113-0033, Japan}
\affiliation{RIKEN Nishina Center, 2-1 Hirosawa, Wako, Saitama 351-0198, Japan}
\affiliation{KU Leuven, Instituut voor Kern- en Stralingsfysica, 3000 Leuven, Belgium}

\author{H. Utsunomiya}\email{hiro@konan-u.ac.jp}
\affiliation{Konan University, Department of Physics, 8-9-1 Okamoto, Higashinada, Kobe 658-8501, Japan}

\date{\today}

\begin{abstract}
 
New measurements of photofission and photoneutron reactions on $^{238}$U and $^{232}$Th in the Giant Dipole Resonance (GDR) energy region have been performed at the laser Compton-scattering $\gamma$-ray source of the NewSUBARU synchrotron radiation facility using a high-and-flat efficiency moderated $^3$He detection array. The neutron-multiplicity sorting of high-multiplicity fission neutron coincidence events has been performed using a dedicated energy dependent, multiple firing statistical treatment. The photoneutron $(\gamma,\,in)$ with $i$~=~1~--~3 and photofission $(\gamma,\,F)$ reactions have been discriminated by considering a Gaussian distribution of prompt-fission-neutron (PFN) multiplicities predicted by the theory of evaporation in sequential neutron emission from excited fission fragments. We report experimental $(\gamma,\,n)$, $(\gamma,\,2n)$, $(\gamma,\,3n)$ and $(\gamma,\,F)$ cross sections, average energies of PFNs and of $(\gamma,\,in)$ photoneutrons, as well as the mean number of PFNs per fission act and the width of the PFNs multiplicity distribution. Based on these primary experimental results and combined with reasonable assumptions, we extract also the first- and second-chance fission contributions. The new experimental results are compared with statistical-model calculations performed with the EMPIRE-3.2 Malta and TALYS-1.964 codes on the present data and with prompt fission emission calculations obtained with the Los Alamos model in the frame of the most probable fragmentation approach with and without sequential emission. 

\end{abstract}

\maketitle


\section{Introduction}

Electromagnetic probes are among the first tools to have been used for investigating the atomic nucleus. Such studies at high $\sim$10--20~MeV incident $\gamma$ energy give insight into the properties of nuclear matter under the extreme conditions of out-of-phase oscillation of the protons against the neutrons, known as the isovector giant dipole resonance~\cite{bracco_tamii_2019,bortignon_bracco_1998,harakeh_2001} and here referred to as GDR. The GDR excitation function reveals important nuclear quantities, such as the quadrupole deformation parameter, the symmetry energy and its energy dependence, the electric dipole polarizability~\cite{goriely_2020}, and the $\gamma$-ray strength function ($\gamma$SF)~\cite{goriely_2019} which, under the Brink hypothesis~\cite{brink_1955,crespo_campo_2018}, is used to describe the $\gamma$-ray cascades in nuclear reactions~\cite{herman_2007_empire,koning_2023_talys}. 

For medium and high atomic number target nuclei, the charged-particle emission from the GDR excited states is highly suppressed by the large Coulomb barrier. Thus, with the exception of few proton rich nuclei, the photoabsorption cross section above the neutron separation energy is well approximated by the sum of $(\gamma,\,inx)$ reactions with $i$ neutrons in the final state, where $i$~=~1~--~4.

For actinides, where the photofission reaction also occurs, additional information related to the shape of the multiple fission barrier and properties of the nucleus in hyper- and super-deformed states can be extracted~\cite{csige_2022}, complementing the picture obtained with hadronic probes~\cite{msin_2021}. For example, considering the limited options in the types of actinide isotopic targets, the properties listed previously can be extracted in photonuclear reactions for nuclei that appear as a second chance in the neutron-induced fission reaction on the same target, being well known that the fission barrier properties can be extracted with better accuracy for the main compound nucleus in the fission chain as compared to the subsequent ones.

Most of the existing GDR photofission and photoneutron cross sections have been measured using quasi-monochromatic $\gamma$-ray beams obtained by positron in flight annihilation at the Saclay~\cite{schuhl_1961_saclay} and Livermore~\cite{hatcher_1961_livermore} facilities. The limited available photon fluxes hindered thin targets experiments with direct fission fragment detection. Instead, high-efficiency neutron detection systems were used for pulsed $\gamma$-ray beam neutron coincidence detection experiments~\cite{Berman_1975_GDR_review}. 

Phenomenological descriptions for the prompt fission neutron multiplicity distribution have been employed in the associated neutron multiplicity sorting techniques~\cite{Fraisse_2023} in order to discriminate neutron contributions from the photoneutron and photofission reactions. The Saclay group made use of prompt fission neutron emission multiplicities extracted from the neutron-induced fission study of Soleilhac \emph{et al.}~\cite{Soleilhac1969}, while the Livermore group used the formalism introduced in Terrell's theory of evaporation in sequential neutron emission from excited fission fragments~\cite{Terrel_1957_PFN}. However, discrepant results have been obtained for the $^{237}$Np, $^{238}$U and $^{232}$Th nuclei measured at the two facilities~\cite{Veyssiere1973,Caldwell1980_PRC,berman_1986}, with both photofission and photoabsorption cross sections systematically higher at Livermore than at Saclay. 

Recently, the IAEA launched a Coordinated Research Project on Photonuclear Data and Photon Strength Functions (Code F41032; 2016-2019)~\cite{goriely_2019,PND2020,Dimitriou15,Kawano2020} which had as one of its main objectives to solve such long-standing discrepancies between Saclay and Livermore data through new photonuclear measurements. In the IAEA-CRP, GDR photoneutron cross sections for 11 nuclei from $^9$Be to $^{209}$Bi~\cite{Gheorghe2017,Kawano2020} were measured at the laser Compton-scattering (LCS) $\gamma$-ray beam line at the NewSUBARU synchrotron radiation facility~\cite{amano_2009,horikawa_2010} of SPring8, Japan. The use of a new high-and-flat efficiency neutron detection system and associated multiplicity sorting techniques \cite{Utsunomiya17,Gheorghe2021} along with low-background and energy-variable quasi-monochromatic LCS $\gamma$-ray beams helped to resolve part of the long-standing discrepancy between the Livermore and Saclay data of partial and total photoneutron cross sections. 

In the present work, we report experimental results of $^{238}$U and $^{232}$Th photoneutron and photofission measurements performed at NewSUBARU following the IAEA CRP. In Sect.~\ref{sec_exp}, the experimental method is described. The data analysis method, consisting in neutron multiplicity sorting and energy unfolding of raw experimental data, as well as the assumptions and procedures applied for discriminating the first- and second-chance fission contributions are described in Sect.~\ref{sec_data_analysis}. Results are discussed and compared with preceding data and theoretical predictions in Sect.~\ref{sect_results}. These include prompt neutron emission results, i.e. mean numbers of prompt-fission-neutrons (PFNs), width of PFN multiplicity distribution and average energies of PFNs, and their comparison with Los Alamos model predictions. Photoneutron and photofission cross sections and average energies of neutrons emitted in $(\gamma,\,in)$ reactions are also compared with statistical model calculations. More details on the statistical model calculation are given in Sect.~\ref{sec_stat_mod_calc}. Finally, a summary is given in Sect.~\ref{sec_summary}.   

\begin{figure*}[t]
\centering
\includegraphics[width=0.65\textwidth, angle=0]{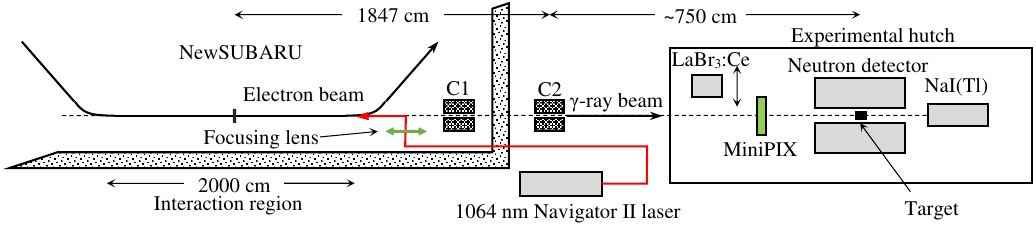} 
\put(-350,0){(a)}
\quad \quad 
\quad \quad 
\includegraphics[width=0.15\textwidth, angle=0]{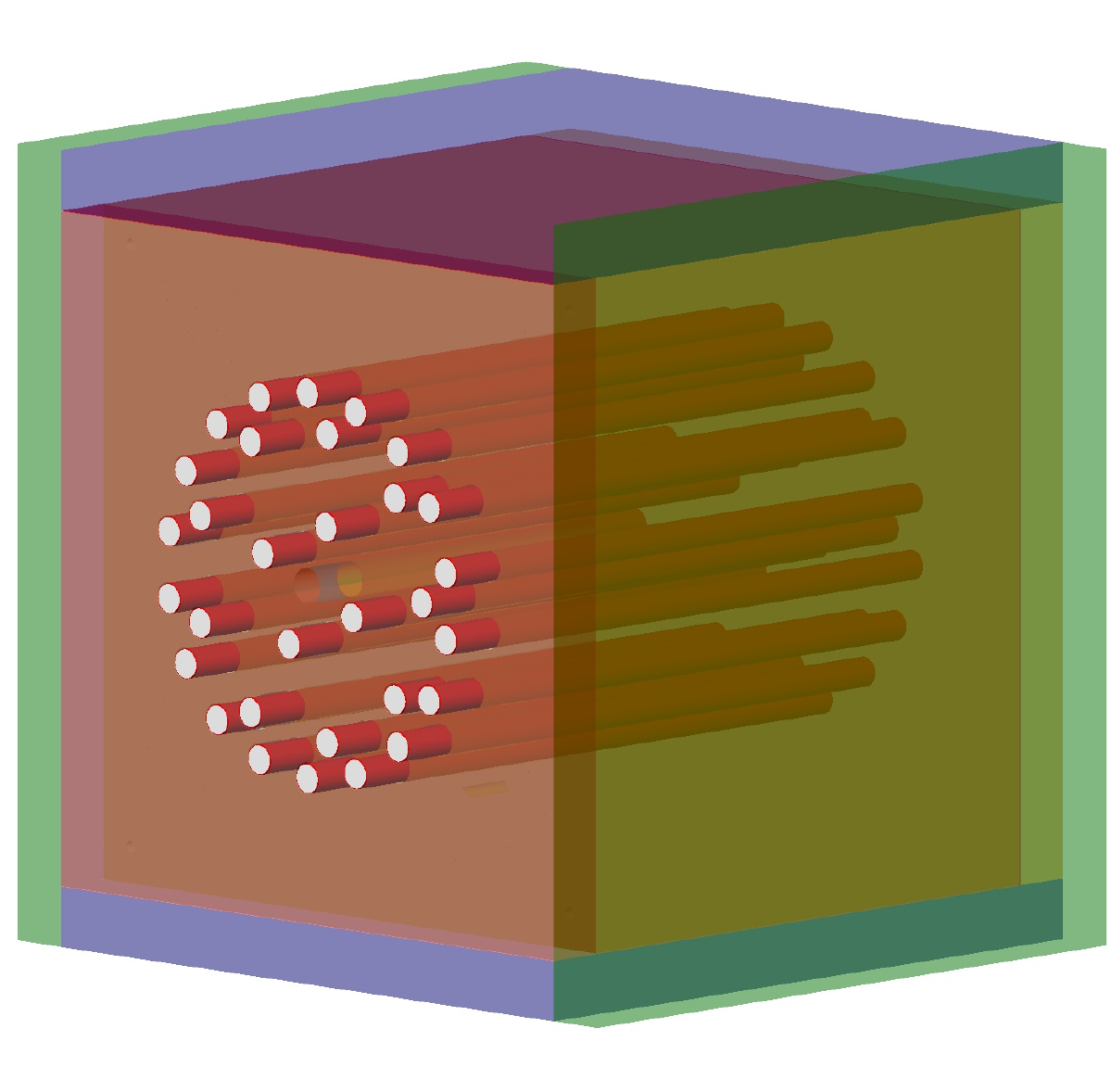} 
\put(-100,0){(b)}
\caption{(a) LCS $\gamma$-ray beam line BL01 and the GACKO experimental hutch at the NewSUBARU synchrotron radiation facility. (b) Diagram showing the $^{3}$He counters layout in the flat efficiency neutron detector (FED).}\label{fig_01_BL01_JINST2023}     
\end{figure*}

\section{Experimental method} 
\label{sec_exp}

The measurements were carried out in the experimental hutch GACKO (Gamma collaboration hutch of Konan University) of the laser-Compton scattering $\gamma$-ray beamline at the NewSUBARU synchrotron radiation facility. A schematic diagram of the experimental setup is shown in Fig.~\ref{fig_01_BL01_JINST2023}(a). The LCS $\gamma$-ray beams irradiated the targets placed at the center of a moderated $^3$He flat-efficiency neutron detection array shown in Fig.~\ref{fig_01_BL01_JINST2023}(b). A large volume NaI detector placed downstream of the flat efficiency neutron detector (FED) was used for inbeam monitoring of the LCS $\gamma$-ray beam flux. The data were recorded in a triggerless list mode, using an eight-parameter 25~MHz digital data acquisition (DAQ) system. The system collected the time and energy signals of the NaI detector, the arrival time of neutrons recorded by the $^3$He counters and the clock signals which triggered the laser beam. Event mode structured data files were constructed offline using the clock signals as time reference. The energy spectra of the incident photon beams were recorded between irradiations by a LaBr$_3$:Ce detector.  

\subsection{Gamma-ray beams}

Quasi-monochromatic $\gamma$-ray beams were generated at energies from 5.87 to 20.14 MeV in the inverse Compton scattering of 1064 nm photons from the Navigator II solid state laser with relativistic electrons in the NewSUBARU storage ring. A system of double 10~cm thick Pb collimators with C1 = 3 mm and C2 = 2 mm diameter was used to limit the $\gamma$-ray beam spot size and define the energy resolution. Electron beam energies were tuned at 40 energy values between 589.89 and 1071.78 MeV. The electron beam energy has been calibrated with the accuracy on the order of 10$^{-5}$ and is reproduced by an automated control of the beam-optics parameters in both deceleration down to 0.5 GeV and acceleration up to 1.5 GeV after every injection of an electron beam at the nominal energy 974 MeV from a linear accelerator \cite{Utsunomiya14}. After injection, the electron beam current slowly dropped from 300~mA to $\sim$100~mA, with a typical beam lifetime of 8~hours.   

The Navigator II laser was operated in Q-switch mode at 1~kHz frequency, corresponding to 1~ms interval between tens of ns wide laser pulses. The NewSUBARU electron beam bunches have a 500~MHz frequency and 60~ps width. Thus, the LCS $\gamma$-ray beam time structure follows the slow laser and the fast electron time structure, with LCS $\gamma$-rays generated in bunches corresponding to each laser light pulse. For background subtraction, the laser had also a slow, 10~Hz frequency pulsed macro-time structure of 80~ms beam-on followed by 20~ms beam-off.  

Given the large number of electron-laser photon interactions and the small Compton scattering cross section, the number of $\gamma$-ray photons in each pulse follows a Poisson probability distribution. During irradiations, the NaI(Tl) detector of 8" diameter and 12" length recorded multiphotons from the same $\gamma$-ray bunch, generating multiphoton (pileup) spectra. The multiphoton spectra were processed through the pile-up/Poisson fitting method \cite{Kondo11,Utsunomiya18} to determine the incident photon flux. During the experiment, the mean number of photons per pulse varied between 5 and 15, corresponding to incident photon fluxes of (4~--~12)~$\cdot$~10$^3$ photons per second. 

\begin{figure}[t]
\centering
\includegraphics[width=0.45\textwidth, angle=0]{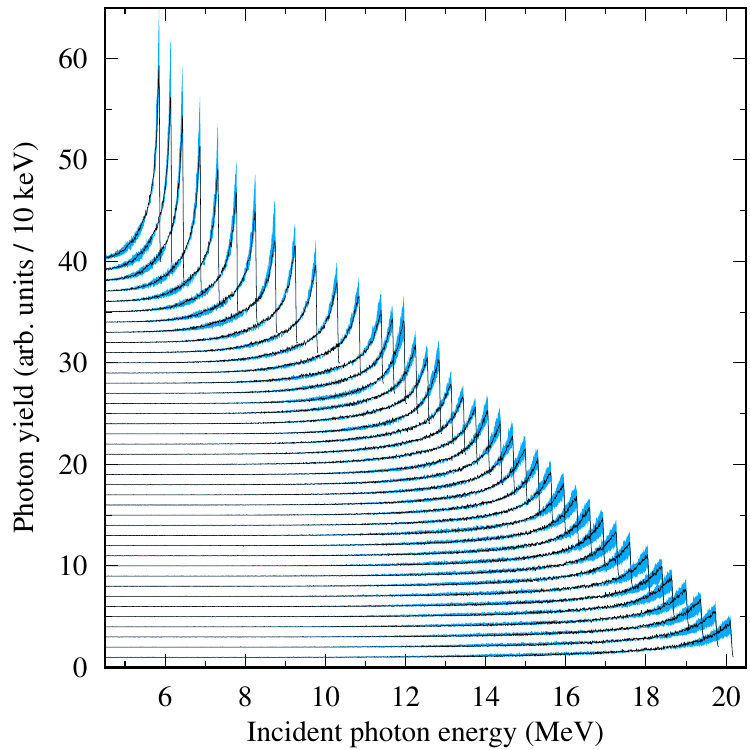} 
\caption{The spectral distributions of the 40 incident LCS $\gamma$-ray beams used in the present experiment, as obtained by Monte Carlo simulations with the \texttt{eliLaBr} code~\cite{eliLaBr_github,filipescu_NIM_LCS_22,filipescu_2022_POL}. The energy spread in full width at half maximum varies between 120~keV and 750~keV for the 5.87 and the 20.14~MeV LCS $\gamma$-ray beams, respectively. The blue band shows the energy spectra uncertainty. Each curve is offset along the vertical axis by 1~arb.~unit for clarity.}\label{fig_inc_spectra_lin2}     
\end{figure}

Single-photon spectra of the LCS $\gamma$-ray beams were measured in between irradiations with a 3.5" diameter $\times$ 4.0" length LaBr$_3$(Ce) detector in a continuous wave (CW) mode of the Navigator II laser and at a reduced laser power. The experimental spectra for each irradiation energy have been reproduced using the dedicated \texttt{eliLaBr} LCS $\gamma$-ray source simulation code \cite{eliLaBr_github,filipescu_NIM_LCS_22,filipescu_2022_POL} implemented using the \textsc{Geant4} package~\cite{geant_allison_2016}. The simulated incident energy distributions of the LCS $\gamma$-ray beams are shown in Fig.~\ref{fig_inc_spectra_lin2}. The central black lines are the results of simulations obtained with parameters that best reproduce the experimental LaBr$_3$:Ce spectra. The blue band shows the uncertainty of the simulation and was obtained using sets of parameters different from the optimal ones but which satisfactorily reproduce the response of the LaBr$_3$:Ce detector. 

\subsection{Targets}

Nuclear fuel materials of 8.62 g ThO$_2$ and 4.06 g U$_3$O$_8$ shielded in pure-aluminum cylindrical containers of 8 mm inner diameters were irradiated with LCS $\gamma$-ray beams. The target and thus the FED alignment to the LCS $\gamma$-ray beam was done by monitoring the visible synchrotron radiation as a guide. Measurements with a MiniPIX X-ray camera~\cite{minipix_website,granja_2022} reproduced by Monte Carlo \texttt{eliLaBr} simulations show that, for the present collimation configuration, the beamspot on target is 4~mm in diameter~\cite{Ariizumi_2023}, which is sufficiently smaller than the 8~mm diameter of the irradiated samples.

An empty Al container was used at energies above $1n$ threshold for $^{27}$Al at 13.06 MeV to measure contributions from Al to neutron events. Contributions from oxygen nuclei to neutron events were measured at energies above neutron threshold at 15.66 MeV for $^{16}$O by using a 10 cm H$_2$O target in an Al cylinder of 14 mm inner diameter with entrance and exit windows of 25.4 $\mu$m Kapton foils. Contributions from the Kapton foils were also checked with an empty cylinder. Table~\ref{tab_targets} lists the properties of the Th, U, Al and H$_2$O targets used in this work. 

The amount of neutron multiplication through neutron-induced fission reactions occurring in the target materials as photoneutrons exit the actinide samples was investigated by \textsc{Geant4.11} neutron transport simulations performed for realistic Maxwell PFNs spectra and typical PFNs mean multiplicities. It follows from the simulation that, due to the small amount of target material, such effects are negligible.

\begin{table}[b]
\caption{\label{tab_targets} Targets used in the experiment: areal density and photon transmission through the samples ($\mathcal{T}$) at the minimum and maximum $\gamma$-ray energies investigated. All targets have natural isotopic abundances.}
\begin{ruledtabular}
\begin{tabular}{lcccc}
   \textrm{Target}                           & \textrm{U$_3$O$_8$} & \textrm{ThO$_2$}& \textrm{Al} & \textrm{H$_2$O}\\ \colrule
    Areal density (g/cm$^2$)                 &  8.08               & 17.15           & 1.08        & 5.38           \\
    $\mathcal{T}_{E_\gamma=6MeV}$ ($\%$)     &  71.5               & 48.5            & 97.2        & 82.2\footnote{for $E_\gamma$~=~16~MeV.}\\
    $\mathcal{T}_{E_\gamma=20MeV}$ ($\%$)    &  64.6               & 38.2            & 97.7        & 83.1           \\
\end{tabular}
\end{ruledtabular}
\end{table}

\subsection{Flat efficiency neutron detector (FED)}

\begin{figure}[t]
\centering
\includegraphics[width=0.45\textwidth, angle=0]{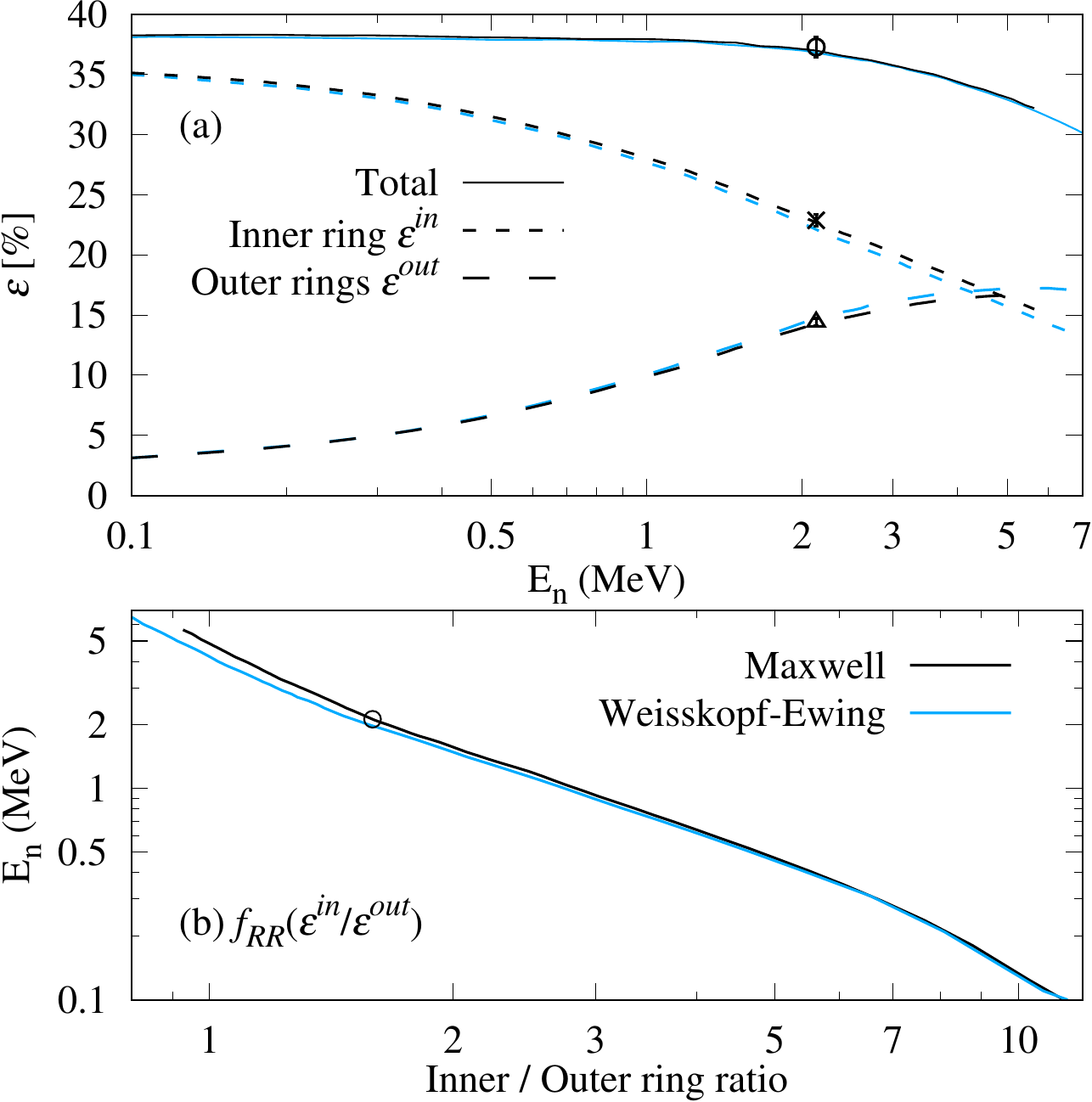} 
\caption{(a) Neutron detection efficiency obtained by MCNP simulations for Maxwell (black) and neutron evaporation spectra (blue) compared with the experimental $^{252}$Cf calibration result. (b) The average neutron energy as function of the inner ring / outer rings detection efficiency ratio.}  \label{fig_fed_eff_new}
\end{figure}

The targets were placed in the center of a flat-efficiency moderated neutron detection array of 31 identical $^3$He counters (10 atm., 2.5 cm diameter $\times$ 45 cm active volume)~\cite{Utsunomiya17}. The array consists of three concentric rings of 4, 9, and 18 $^3$He counters embedded in a polyethylene moderator at 5.5, 13.0 and respectively 16.0 cm from the $\gamma$-ray beam axis, as shown in Fig.~\ref{fig_01_BL01_JINST2023}(b). 

Fig.~\ref{fig_fed_eff_new}(a) shows MCNP simulations for the total neutron detection efficiency along with efficiencies for the inner ring of counters and for the sum of the two outer rings. In the simulations, we considered Maxwell neutron spectra characteristic for PFNs (black lines), as well as neutron evaporation spectra described by the Weisskopf-Ewing function~\cite{Weisskopf_1937} (blue lines). The simulation results are represented at the corresponding average neutron emission energies. An experimental calibration with a $^{252}$Cf source of known activity is also shown in Fig.~\ref{fig_fed_eff_new}(a). The total detection efficiency varies within 5$\%$ from 38.1$\%$ (37.8$\%$) at 10~keV to 33.1$\%$ (32.9$\%$) at 5~MeV for Maxwell (evaporation) neutron spectra. The good agreement between the two curves demonstrates a robust flatness of the total neutron detection efficiency, and thus a good insensitivity of the extracted cross sections as a function of the specific neutron emission spectrum. 

Instead, the partial detection efficiency of the inner ring and that of the summed two outer rings depend on the average neutron energy. Fig.~\ref{fig_fed_eff_new}(b) shows that the $f_{RR}$ ratio between the two of them decreases with the neutron energy, a feature which is used through the ring ratio (RR) method to extract the average neutron energy. As described in~Refs.~\cite{Berman_1975_GDR_review,Gheorghe2021}, the neutron energy is obtained by evaluating the simulated $f_{RR}$ function at the experimental ratio between the number of neutrons recorded in the inner and respectively outer rings of counters. We note that, unlike the total detection efficiency, the $f_{RR}$ functions computed for Maxwell and neutron evaporation spectra diverge for neutron energies above $\sim$1~MeV, which is the region of interest for PFNs average energies. This indicates that the RR extracted average neutron energies are in fact sensitive to the choice of the simulated neutron emission spectra. 

Thus, the accuracy of the efficiency simulations performed by modeling the PFN spectra by Maxwell functions was tested using dedicated predictions for PFN spectra emitted in the photofission reactions on $^{232}$Th and $^{238}$U~\cite{Tudora_2023_ND}. Predictions of PFNs spectra and multiplicities were obtained through calculations in the frame of the most probable fragmentation approach with the Los Alamos (LA) model without~\cite{madland_2017} and with~\cite{tudora_2020a} sequential emission, using input parameters provided by the recent systematic of Ref.~\cite{tudora_2020b} and fission chance probabilities based on present EMPIRE statistical model calculations. Throughout this work, we will refer to them as LA model predictions. Good agreement was obtained between the simulations performed by sampling neutrons from the so obtained spectra predictions and by using Maxwell spectra. 

\begin{figure}[t]
\centering
\includegraphics[width=0.49\textwidth, angle=0]{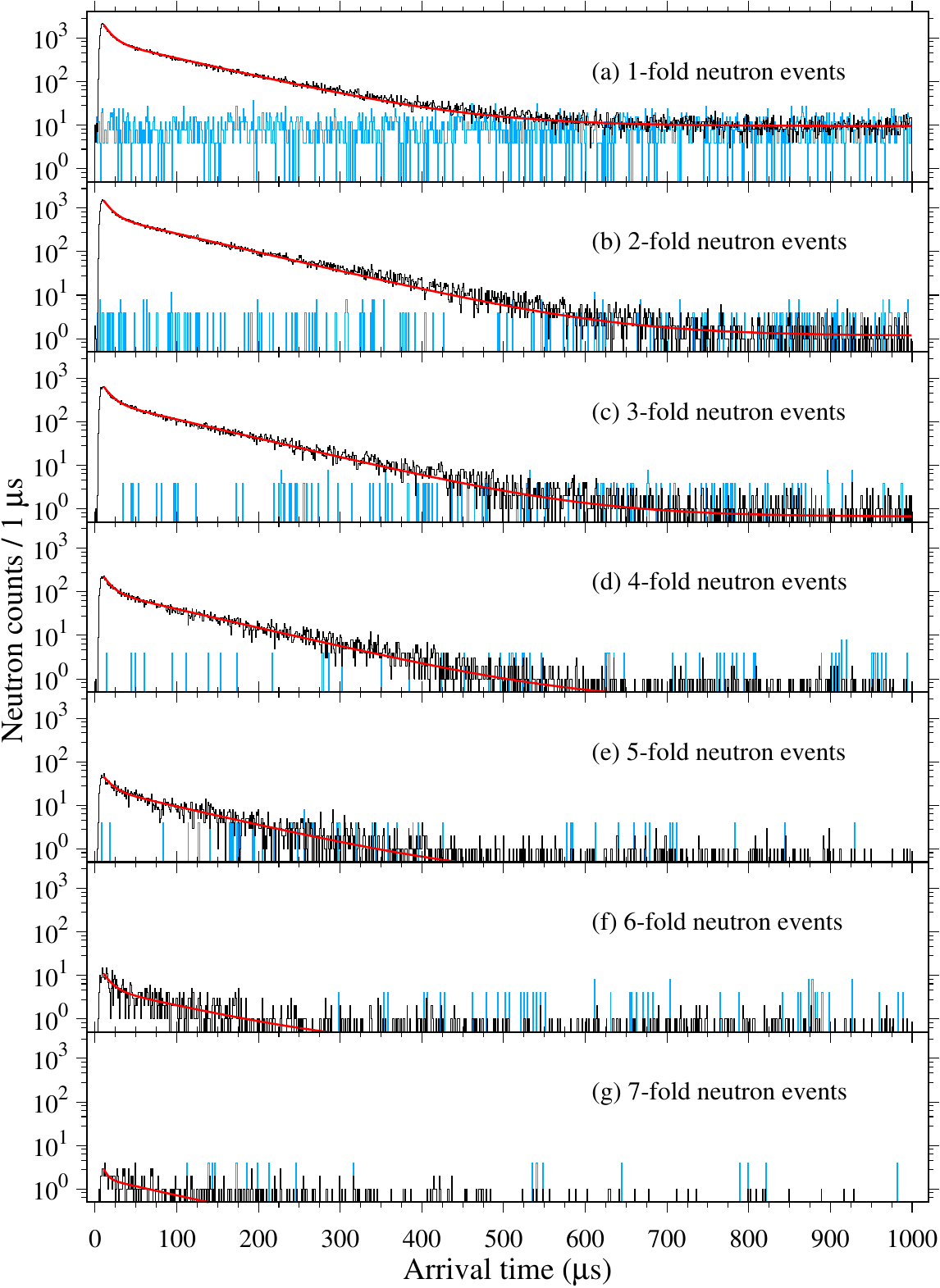} 
\caption{Arrival-time distributions of neutrons recorded by the FED in the photoneutron and photofission reactions on $^{238}$U at 18.67~MeV. 
The experimental neutron counts recorded during beam-on (black) and beam-off (blue) are displayed for (a) 1-fold, (b) 2-fold, (c) 3-fold, (d) 4-fold, (e) 5-fold, (f) 6-fold and (g) 7-fold neutron events. The red lines show least square fits to the experimental distributions obtained using a sum of two exponentials and a constant background component.}  \label{fig_neutron_time}
\end{figure}

\begin{figure}[t]
\centering
\includegraphics[width=0.49\textwidth, angle=0]{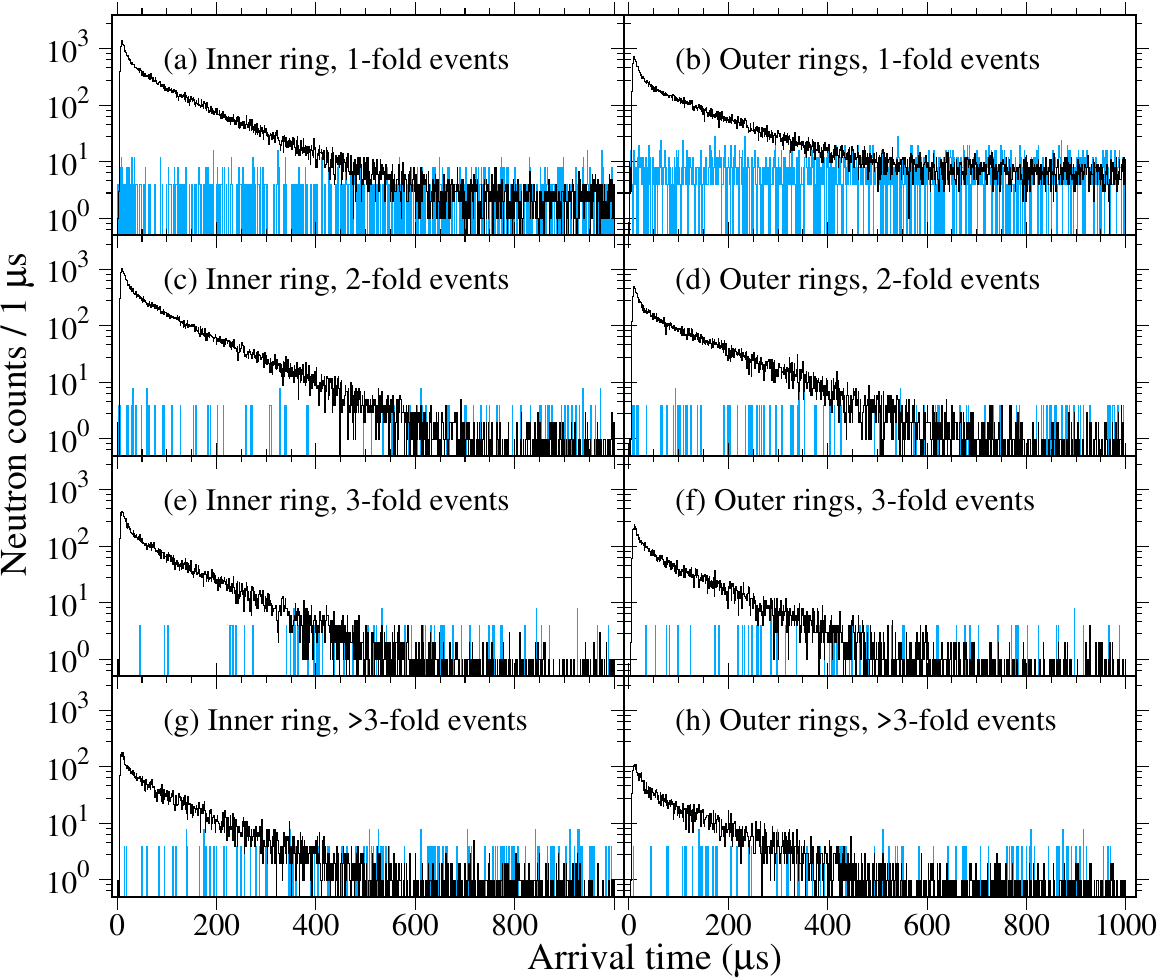} 
\caption{Same as Fig.~\ref{fig_neutron_time}, where the experimental neutron counts recorded by the inner ring counters (left column) and by the counters in the two outer rings (right column) are displayed for (a, b) 1-fold, (c, d) 2-fold, (e, f) 3-fold and (g, h) 4- to 7-fold neutron events.}  \label{fig_ring_time}
\end{figure}

The radial arrangement of the $^3$He counters allowed the investigation of the azimuthal asymmetry of PFNs emission through individual scaller monitoring of the detection rate per each counter. Such measurements have recently been made at the HI$\gamma$S LCS $\gamma$-ray source to extract polarization asymmetries in subbarrier fission~\cite{Mueller_2014,Silano18}. However, azimuthal asymmetries haven't been observed in the present GDR measurements due to the cumulative effect of the overlap between PFNs and $(\gamma,\,xn)$ photoneutron ones, and the fact that the number of possible values of the (J,$\Pi$) quantum numbers characterizing transition states increases with the increase in the excitation energy.

\subsection{Multi-neutron coincidence data}

For the actinide $^{232}$Th and $^{238}$U targets irradiated in the present experiment, the neutrons were emitted in both photoneutron and photofission reactions. In Table~\ref{table_isotopes}, we give neutron emission thresholds values and, as a reference, we also list the B$_F(\gamma,f)$ first- and B$_F(\gamma,nf)$ second-chance fission thresholds extracted from preceding photofission and neutron induced fission experiments (see Ref.~\cite{Caldwell1980_PRC} and references therein). The first- and second-chance fission thresholds determined in the present experiment (see Sect.~\ref{sec_res_separation_fis_chance}) agree, within the experimental limits, with those given in Table~\ref{table_isotopes}. 

While the $\texttt{x}$ maximum neutron emission multiplicity in photoneutron reactions varied between 0 and 3, depending on the incident photon energy and the characteristic $i$ neutron separation energies $S_{in}$, the photofission reactions emit up to $\sim$9 PFNs per fission act. The primary experimental information from which the competing $(\gamma,\,in)$ and $(\gamma,\,F)$ cross sections are extracted through neutron-multiplicity sorting are the neutron coincidence events. 

Here, an event is defined as an $i$-fold coincidence when $i$ neutrons are recorded in the 1 ms interval between two consecutive $\gamma$ pulses. The 1~ms interval was chosen based on the neutron die-away time inside the FED~\cite{Utsunomiya17,Gheorghe2017,Gheorghe2021}. We note that, in \emph{single-firing} conditions, no more than one nuclear reaction is induced by each photon beam pulse, and thus all recorded neutron coincidence events with multiplicities greater than $\texttt{x}$ can be assigned to photofission reactions. In the present experiment, the incidence of \emph{multiple-firing} events has been reasonably minimized, however not eliminated, by using $\gamma$-ray beams of limited 5~to~15 mean photon multiplicities per bunch.  

Figure~\ref{fig_neutron_time} shows the histogram of the neutron arrival time for (a) one-, (b) two-, $\dots$, (g) seven-fold coincidence events in the photon induced reactions on $^{238}$U at 18.67~MeV, above the $S_{3n}$~=~17.82~MeV. 
The histograms were built for an irradiation of 2 hours, the typical duration of a measurement in this experiment. The black histograms correspond to the beam-on data and the blues ones to the beam-off data normalized by a factor of 4~=~80~ms~/~20~ms. The background subtraction procedure, which relies on fitting the time distribution with a sum of exponentials plus a flat background, has been discussed in Refs.~\cite{Utsunomiya17,Gheorghe2017,Gheorghe2021}. We note that the background component derived from the fitting procedure reproduces the experimental beam-off background level.

Multi-neutron coincidence events were also discriminated by the firing ring in order to apply the ring ratio method and determine the average energies of neutrons detected in $i$-fold coincidences. Figure~\ref{fig_ring_time} shows the histogram of the neutron detection time by the inner ring and, respectively by the summed two outer rings in (a,b) one-, (c,d) two-, (e,f) three and (g,h) $>$3 neutron coincidence events for 18.67 MeV $\gamma$-ray beam incident on $^{238}$U. Considering that $E_\gamma$~$<$~$S_{4n}$ and the experiment was conducted in close to \emph{single-firing} conditions, a good approximation for the average energy of the total PFN spectrum can be determined by applying the RR-method on the summed up ring data for multiplicities higher than 3 shown in Fig.~\ref{fig_ring_time}(g,h). Thus, for each incident energy, we have summed up the arrival time histograms in the inner and respectively outer rings for neutrons recorded in events of multiplicities higher than the maximum $(\gamma,\,in)$ photoneutron multiplicity $\texttt{x}$. 

\begin{table}[b]
\caption{\label{table_isotopes} Separation energies for one ($S_n$), two ($S_{2n}$) and three ($S_{3n}$) neutrons and B$_F(\gamma,f)$ first- and B$_F(\gamma,nf)$ second-chance fission thresholds for $^{232}$Th and $^{238}$U. The B$_F(\gamma,f)$ is the fission barrier for the main compound nucleus and the B$_F(\gamma,nf)$ values are obtained by adding $S_n$ to the B$_F(\gamma,f)$ for the (N-1) nucleus \footnote{all values are given in MeV with $\pm$ 0.2 MeV uncertainty.}.}
\begin{ruledtabular}
\begin{tabular}{lccccc}
   \textrm{Target}       & \textrm{S$_n$} & \textrm{S$_{2n}$}& \textrm{S$_{3n}$} & B$_F(\gamma,f)$ & B$_F(\gamma,nf)$ \\ \colrule
    \textrm{$^{232}$Th}  &  6.44          & 11.56            & 18.35             & 6.0             & 12.6 \\
    \textrm{$^{238}$U}   &  6.15          & 11.28            & 17.82             & 5.8             & 12.3           \\
\end{tabular}
\end{ruledtabular}
\end{table}

\section{Data reduction and analysis}
\label{sec_data_analysis}

\subsection{Key experimental observables: $i$-fold neutron cross sections and average energies} \label{subsec_Ni_Ei}

\begin{figure*}[t]
\includegraphics[width=0.98\textwidth, angle=0]{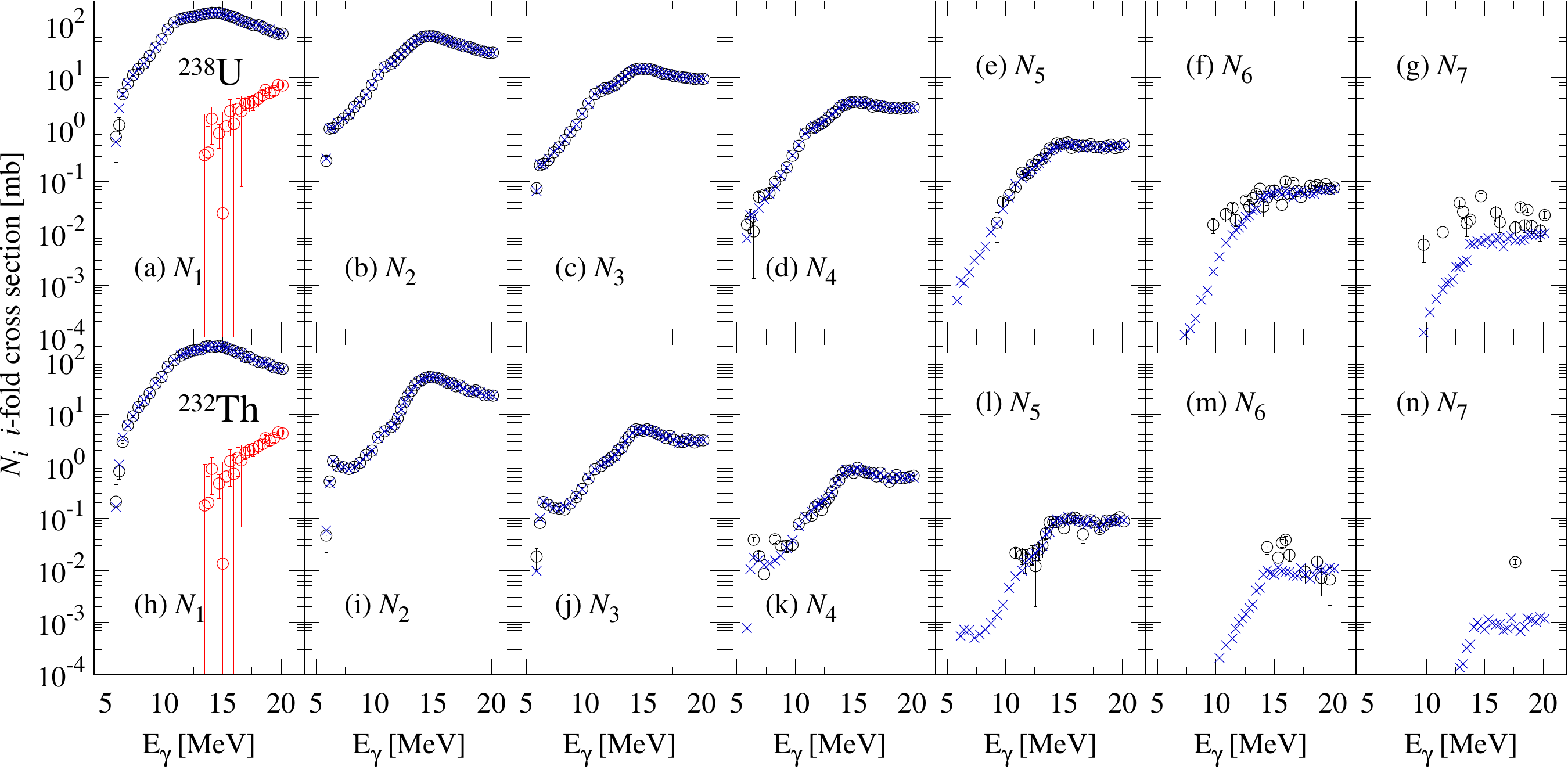} 
\caption{Experimental ($N_i$, empty black dots) and best fit ($N_i^{MF}$, blue crosses) $i$-fold cross sections, defined in Eq.~\eqref{EQ_Nk_rate} as the number of $i$-fold neutron events recorded per incident photon and target nucleus, for the photoneutron and photofission reactions on $^{238}$U (a~--~g) and $^{232}$Th (h~--~n). The empty red dots show the $N_1$ total subtracted background contribution from the Al container and from the Oxygen nuclei in the U$_3$O$_8$ and the ThO$_2$ molecules. The error bars are statistical only. \label{fig_nev_rate}}
\end{figure*}
\paragraph{$N_i$ $i$-fold neutron cross sections.} We define the $N_i$, expressed in cross section units (mb), as the number of $i$-fold neutron coincidence events recorded per incident photon and target nucleus:
\begin{equation} \label{EQ_Nk_rate}
N_i = \cfrac{ \sum_t n_i[t]/i }{N_\gamma n_T \xi} 
\end{equation} 
where $n_i[t]$ is the background subtracted arrival time histogram for neutrons recorded in $i$-fold events. $n_T$ is the concentration of target nuclei, $N_\gamma$ is the incident photon number for the total irradiation time and $\xi=[1-\mathrm{exp}(-\mu L)]/\mu$ is a thick target correction factor given by the target thickness $L$ and attenuation coefficient $\mu$. 

Figure~\ref{fig_nev_rate} shows the experimental (black empty dots) $i$-fold neutron cross sections for (a-g)~$^{238}$U and for (h-n)~$^{232}$Th. From the analysis of the empty Al container and H$_2$O target data we found that small single neutron contributions had to be subtracted from the U$_3$O$_8$ and the ThO$_2$ data, following the normalization procedure described in Appendix~\ref{annex_B}. The values displayed by red empty dots in Fig.~\ref{fig_nev_rate} represent the total subtracted contribution for each experimental $1$-fold point. Although the Al and Oxygen contributions were low, we were able to extract the $^{27}$Al$(\gamma,\,n)$ and the $^{16}$O$(\gamma,\,n)$ cross sections from the empty Al container and H$_2$O target data, as shown in Appendix~\ref{annex_B}. 

Because of the non-unity detection efficiency of $\sim$37$\%$, we notice in Fig.~\ref{fig_nev_rate} that the counting statistics of high multiplicity neutron coincidence events becomes increasingly poor with increasing neutron multiplicity. Depending on the $(\gamma,\,F)$ reaction cross section and PFN multiplicity distribution, the maximum neutron multiplicities observed experimentally in the present study varied between 3 and 7. Table~\ref{table_nev_summary} summarizes the maximum recorded neutron multiplicities, as observed in Fig.~\ref{fig_nev_rate}. 

In general, we note that the maximum recorded multiplicity at the same excitation energy is lower for $^{232}$Th than for $^{238}$U. In particular, for $^{232}$Th, the two lowest energy points, at 5.87 and 6.16~MeV excitation energy, have a maximum recorded neutron multiplicity of 3, which is the minimum value obtained in the present experiment. However, these two points are below the $S_n$, and in fact all recorded neutrons originate from $(\gamma,\,F)$ reactions only, which significantly simplifies the neutron-multiplicity sorting procedure and lowers the requirements on the observed neutron multiplicity, as described in Appendix~\ref{annex_A}. 

\begin{table}[b]
\caption{\label{table_nev_summary} Maximum neutron multiplicity $\texttt{N}^*$ observed for $^{232}$Th and $^{238}$U for energy regions defined by the $S_{\texttt{x}n}$ separation energies in each isotope, where $\texttt{x}$~=~1 to 3. Most common values are given together with exceptions listed in brackets, as a summary of the data shown in Fig.~\ref{fig_nev_rate}. } 
\begin{ruledtabular}
\begin{tabular}{lcccc}
                     & $E_{\gamma}\!<\!S_n$ & $S_n\!<\!E_{\gamma}\!<\!S_{2n}$& $S_{2n}\!<\!E_{\gamma}\!<\!S_{3n}$ & $S_{3n}\!<\!E_{\gamma}$ \\ \colrule
$\texttt{N}^*_{^{232}\textrm{Th}}$          &  3                   & 4, (3, 5)                      & 5, (4, 6, 7)                       & 5, 6     \\
$\texttt{N}^*_{^{238}\textrm{U}}$           &  4                   & 4, 5, (6, 7)                   & 6, 7, (5)                          & 7, (6)  \\
$\texttt{x}\!\!+\!\!3$  &  3                   & 4                              & 5                                  & 6       \\
\end{tabular}
\end{ruledtabular}
\end{table}

\begin{figure*}[t]
\includegraphics[width=0.98\textwidth, angle=0]{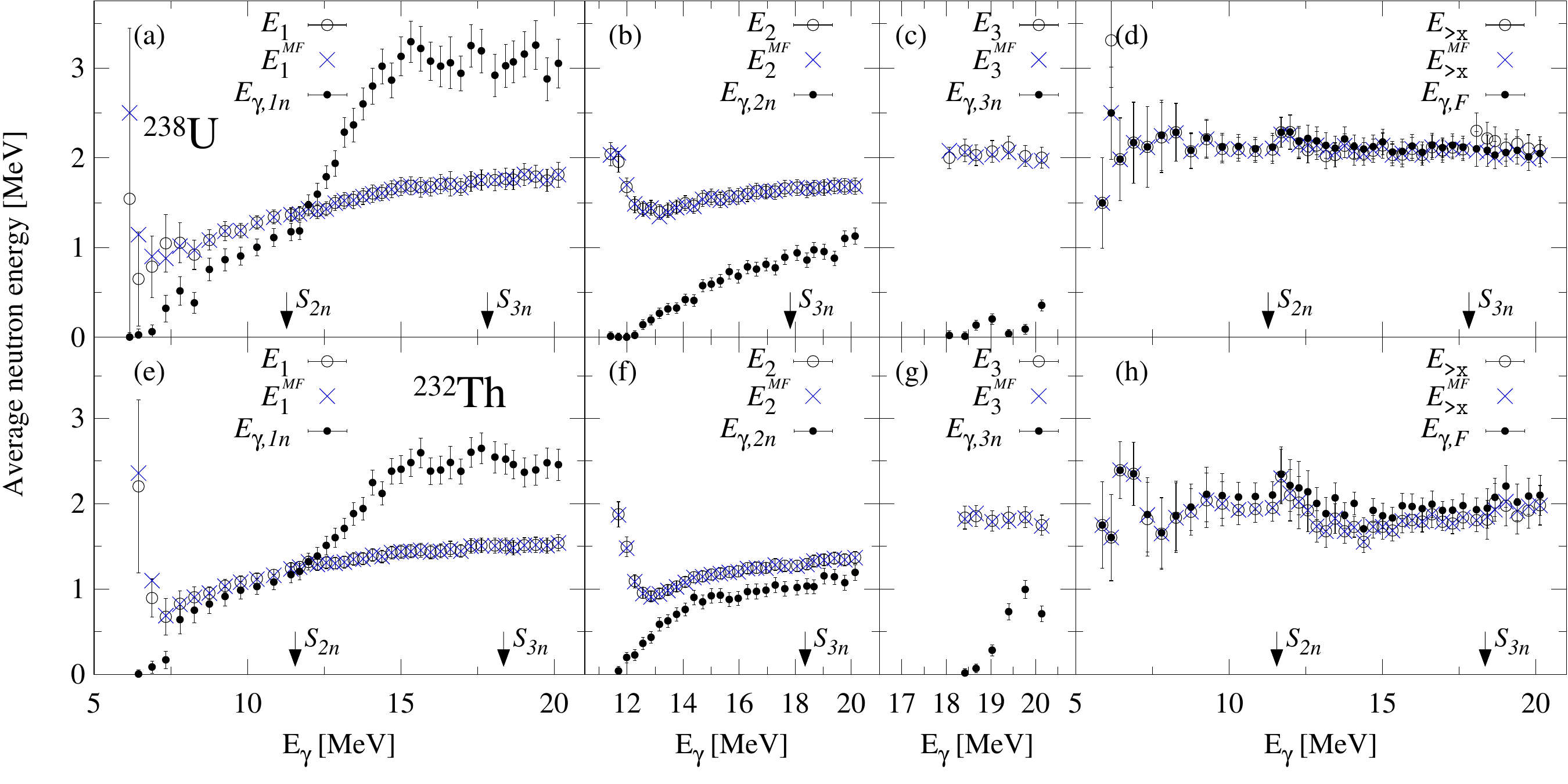} 
\caption{Average energies in monochromatic approximation of photoneutron and PFNs for (a~--~d) $^{238}$U and (e~--~h) $^{232}$Th: average energies of neutrons recorded in $i$-fold coincidence events (experimental - empty black dots, best fit - blue crosses), multiple firing neutron multiplicity sorting results for the $E_{\gamma,\,xn}$ average energies of $(\gamma,\,xn)$ photoneutrons and $E_{\gamma,F}$ of PFNs (full black dots). The error bars for the average energies of $i$-fold coincidence events represent the statistical component and a 3$\%$ systematic component accounting for the uncertainty in the neutron detection efficiency calibration.  \label{fig_energy_mono}} 
\end{figure*}
\paragraph{$E_i$ $i$-fold average neutron energies.} $E_i$ is the average energy of neutrons recorded in $i$-fold events and has been experimentally determined by the RR method as: 
\begin{equation} \label{EQ_Ei_exp}
E_i = f_{RR}(\sum_t n^{in}_i[t] / \sum_t n^{out}_i[t]),
\end{equation}
where $n^{in}_i[t]$ and $n^{out}_i[t]$ are the background subtracted arrival time histograms for neutrons detected in $i$-fold events by the inner ring and by the summed two outer rings, respectively. 

Figure~\ref{fig_energy_mono} shows the experimental (black empty dots) $E_i$ average energies of neutrons recorded in (a,e)~1-, (b,f)~2-, (c,g)~3- and (d,h)~$>\!\!\texttt{x}$-fold coincidence events for (top)~$^{238}$U and (bottom)~$^{232}$Th, where $\texttt{x}$ is again the maximum $(\gamma,\,in)$ photoneutron multiplicity for each incident energy. Thus, $E_1$, $E_2$ and $E_3$ have been determined only for incident photon energies above $S_n$, $S_{2n}$ and respectively $S_{3n}$. For example, for the two lowest $^{232}$Th points which are below $S_n$, $\texttt{x}$~=~0 and thus the RR-method has been applied on the total inner and respectively outer ring data to obtain the average energy of the PFN spectrum. 

\subsection{Neutron multiplicity sorting} \label{sec_NMS}

Considering a photon beam of $E_\gamma$ energy incident on an actinide target, where S$_{\texttt{x}n}$~$<$~E$_\gamma$~$<$~S$_{(\texttt{x}+1)n}$, the following competitive reactions can be induced:
\begin{itemize}
\item photoneutron ($\gamma,\,in$) reactions of $\sigma_{\gamma,\,in}$ cross sections and $E_{\gamma,in}$ average neutron emission energies, where $i$ takes values from 1 to $\texttt{x}$;
\item photofission $(\gamma,\,F)$ reactions with emission of $i$ PFNs of $E_{\gamma,F}$ average energy and described by a $\rho_i$ multiplicity distribution for which the $\sum_i \rho_i$~=~1 condition is generally reasonably met for maximum PFN emission multiplicities $\texttt{N}$~$\approx$~8$\sim$9. We can express the total photofission cross section $\sigma_{\gamma,\,F}$ in terms of the partial cross sections $\sigma_{\gamma,\,fin}$ for photofission with emission of $i$ PFNs:
\begin{equation}
\sigma_{\gamma,\,F} = \sum_{i=0}^{\texttt{N}} \sigma_{\gamma,\,fin}. 
\end{equation}
\end{itemize}

In order to extract the contributing photoneutron and photofission reactions cross sections and average neutron energies, a neutron multiplicity sorting procedure must be applied on the direct experimental observables $i$-fold cross sections $N_i$ and average energies $E_i$, which were discussed in the previous section. By considering scenarios of increasingly realistic experimental conditions, we here define the problem and introduce the necessary assumptions for solving it:

\paragraph{} Assuming ideal \emph{single-firing} conditions and unity neutron detection efficiency $\varepsilon$, one would experimentally determine the $N_1$ to $N_\texttt{N}$ $i$-fold cross sections, which would directly relate to the photoneutron and photofission cross sections as: 
\begin{equation} \label{eq_ref_ideal_eq_sytem}
N_i  = \sigma_{\gamma,\,in} + \sigma_{\gamma,\,fin}, 
\end{equation}
where of course the $\sigma_{\gamma,\,in}$ photoneutron cross sections are zero for $i$~$>$~$\texttt{x}$. As one would have an underdetermined system of $\texttt{N}$ equations and $\texttt{x}+\texttt{N}$ variables, it wouldn't be possible to discriminate the $\sigma_{\gamma,\,in}$ and $\sigma_{\gamma,\,fin}$ cross sections for low emission multiplicities up to $i$~=~$\texttt{x}$. 

In order to limit the number of independent variables, we made use of Terrell's theory of evaporation in sequential neutron emission from excited fission fragments~\cite{Terrel_1957_PFN} employed also by the Livermore group~\cite{Caldwell_1975_NSE}, and which demonstrates that the $\rho_i$ multiplicity distribution of PFNs can be approximated with a Gaussian function:
\begin{equation}
\sum_{j=0}^{i} \rho_{j} = \cfrac{1}{2} + \cfrac{1}{2} \, f \Big( \cfrac{i - \overline{\nu}_p + 1/2 + b}{\sigma}\, \Big),
\end{equation}
with
\begin{equation}
f(v) = \cfrac{1}{\sqrt{2\pi}} \int_{-v}^v \exp \Big(\cfrac{-t^2}{2} \Big) \, dt
\end{equation}
and
\begin{equation}
b \lesssim 0.01 \approx \cfrac{1}{2} - \cfrac{1}{2} f \, \Big ( \cfrac{\overline{\nu}_p + 1/2}{\sigma} \,  \Big).
\end{equation}
This reduces the number of photofission related variables from $\texttt{N}\approx$~8$\sim$9 to only three independent parameters: the total $\sigma_{\gamma,\,F}$ cross section, the  mean number of PFNs per fission act $\overline{\nu}_p$ and a width parameter $\sigma$:
\begin{equation}\label{eq_cs_gfin_cs_gF}
\sigma_{\gamma,\,fin} = \sigma_{\gamma,\,F}\cdot \rho_i(\overline{\nu}_p, \sigma).
\end{equation}
Thus, in ideal conditions of unit detection efficiency $\varepsilon$, Eqs.~\eqref{eq_ref_ideal_eq_sytem} would be an overdetermined system of $\texttt{N}$ equations and $\texttt{x}$+3 variables. 

\paragraph{} Still in \emph{single-firing} conditions, but considering a realistic energy independent non-unity $\varepsilon$ neutron detection efficiency, the experimental $N_i$ $i$-fold cross sections are now expressed as:
\begin{equation} \label{eq_ref_SF_eq_system}
N_i = \sum_{x=i}^{\texttt{N}} ( \sigma_{\gamma,\,xn} + \sigma_{\gamma,\,fxn} ) \cdot {}_{x}C_{i} \varepsilon^i (1-\varepsilon)^{x-i},
\end{equation}
where ${}_{x}C_{i}$ are the binomial coefficients and, again, for simplicity of notation, we cycled $\sigma_{\gamma,\,xn}$ up to $\texttt{N}$ with zero cross section values for $x$~$>$~$\texttt{x}$.  

However, as shown in Section~\ref{subsec_Ni_Ei}, because of the non-unity detection efficiency of $\sim$37$\%$, the $\texttt{N}^*$ maximum neutron multiplicities observed experimentally have been generally lower than the expected highest PFN emission multiplicity $\texttt{N}$~=~8$\sim$9. We notice in Table~\ref{table_nev_summary} that, for $^{238}$U, the number of experimentally observed multiplicities $\texttt{N}^*$ is either equal to or slightly greater than the $\texttt{x}$+3 number of unknown parameters. For $^{232}$Th however, as lower maximum multiplicities have been experimentally observed, $\texttt{N}^*$ is sometimes lower than $\texttt{x}$+3. In order to account for the limited statistics in registering high-multiplicity neutron events, we found it necessary to constrain the $\sigma$ width parameter of the PFNs multiplicity distribution, which is generally known to show a slow average increase with the excitation energy, both for photon- and neutron-induced fission. As shown in Sec.~\ref{sec_res_PFN_mult}, the width parameter was constrained to the linearly dependent average of the present $\sigma$ experimental results. 

In \emph{single-firing} and flat-efficiency conditions, we can also express the average energies of neutrons recorded in $i$-fold coincidences as:
\begin{equation} \label{eq_ref_ideal_eq_sytem_Ei}
E_i  =\sum_{x=i}^{\texttt{N}} \Big(E_{\gamma,xn} \sigma_{\gamma,\,xn} + E_{\gamma,F} \sigma_{\gamma,\,fxn} \Big ) {}_{x}C_{i} \varepsilon^i (1-\varepsilon)^{x-i} /N_i. 
\end{equation}
By averaging the above expressions for $i>\texttt{x}$, one obtains the natural result that the average PFN energy $E_{\gamma,F}$ is equal to the average energies of neutrons recorded in multiplicities higher than the maximum photoneutron emission multiplicity $\texttt{x}$: $E_{\gamma,F}=E_{>\texttt{x}}$. Thus, the system of $\texttt{x}+1$ equations~\eqref{eq_ref_ideal_eq_sytem_Ei} can be solved to obtain the \emph{single-firing} approximation values for the $E_{\gamma,in}$ and $E_{\gamma,F}$ energies.  

\paragraph{} Moving closer to reality, the situation further complicates by considering the small probability of more than one reaction being induced in the target by the same photon bunch. Following probabilities given by partial cross sections, target areal density and number of incident photons per pulse, all combinations of energetically available reactions can be induced by each beam pulse. Thus, one can no longer assume that all coincident neutrons recorded in a given 1~ms event are originated from the same reaction and, for example, one can no longer assign all neutron events of multiplicities higher than $\texttt{x}$ to photofission reactions. Thus, we have extended the $multiple$-firing statistical method~\cite{Gheorghe2021} originally developed for photoneutron reactions only, by additionally implementing the photofission reaction channels. Although the addition of $(\gamma,\,fxn)$ reaction channels is quite straightforward, we give the complete method in Appendix~\ref{annex_A}. 

In the procedure, $\texttt{N}$ was equal to 9, the highest significant PFNs emission multiplicity which verifies the $\sum_{x=1}^{\texttt{N}}\rho_x(\overline{\nu}_p,\sigma)=1$ condition. We considered $(r_xf_x)$ combinations of $(\gamma,\,xn)$ and $(\gamma,\,fxn)$ reactions, as defined in Eq.~\eqref{eq_ref_rx_fx}, where each individual reaction could be induced for a maximum of 2 times in a given combination, while the maximum total number of reactions induced in each combination was 3. Using the notations given in Appendix~\ref{annex_A}, the $r_x$ and $f_x$ indices in Eq.~\eqref{eq_ref_rx_fx} cycled from 0 to 2, while their sum $r$ given in Eq.~\eqref{eq_ref_def_r_total} took values up to 3. 

We used the $\texttt{CERN}$ $\texttt{minuit}$ package to perform a $\chi^2$ minimization procedure and determine the set of input parameters $\sigma_{\gamma,\,in}$, $E_{\gamma,\,in}$, $\sigma_{\gamma,\,F}$, $\overline{\nu}_p$, $\sigma$ and $E_{\gamma,\,F}$ for which the calculated $i$-fold $N^{MF}_i$ cross sections and $E^{MF}_i$ energies reproduced best the experimental $N_i$ and $E_i$ values. We defined the $\chi^2$ as:   
\begin{equation} \label{label_eq_chi2_MF}
\chi^2 = \sum_{i=1}^{\texttt{N}^*} \cfrac{ (N_i - N^{MF}_i )^2 }{\sigma_{N_i}^2} + \sum_{i=1}^{\texttt{x}+1} \cfrac{ (E_i - E^{MF}_i )^2 }{\sigma_{E_i}^2},
\end{equation}
A preliminary, \emph{single-firing} and energy independent minimization procedure is performed to obtain starting values for the input parameters in the minimization procedure. 

The best-fit $N^{MF}_i$ calculated $i$-fold cross sections (blue crosses) are shown in Fig.~\ref{fig_nev_rate} in comparison with the experimental ones. The cross sections for the $i$-fold multiplicities $N_1$ to $N_6$ are well reproduced by the minimization procedure. Instead, we notice that calculations underestimate the experimental $N_7$ values, especially at excitation energies below $\sim$15~MeV. The best-fit $E^{MF}_i$ calculated $i$-fold average neutron energies (blue crosses) shown in Fig.~\ref{fig_energy_mono} reproduce well the experimental values on the entire excitation energy range and for all neutron multiplicities. We notice the sharp energy drop in $E_2$ at $S_{2n}$ for both $^{238}$U and $^{232}$Th, given by the low neutron energy contribution of the newly opened $(\gamma,\,2n)$ channel.

\subsection{Energy unfolding}\label{subsec_energy_unfold}

\begin{figure}[t]
\includegraphics[width=0.49\textwidth, angle=0]{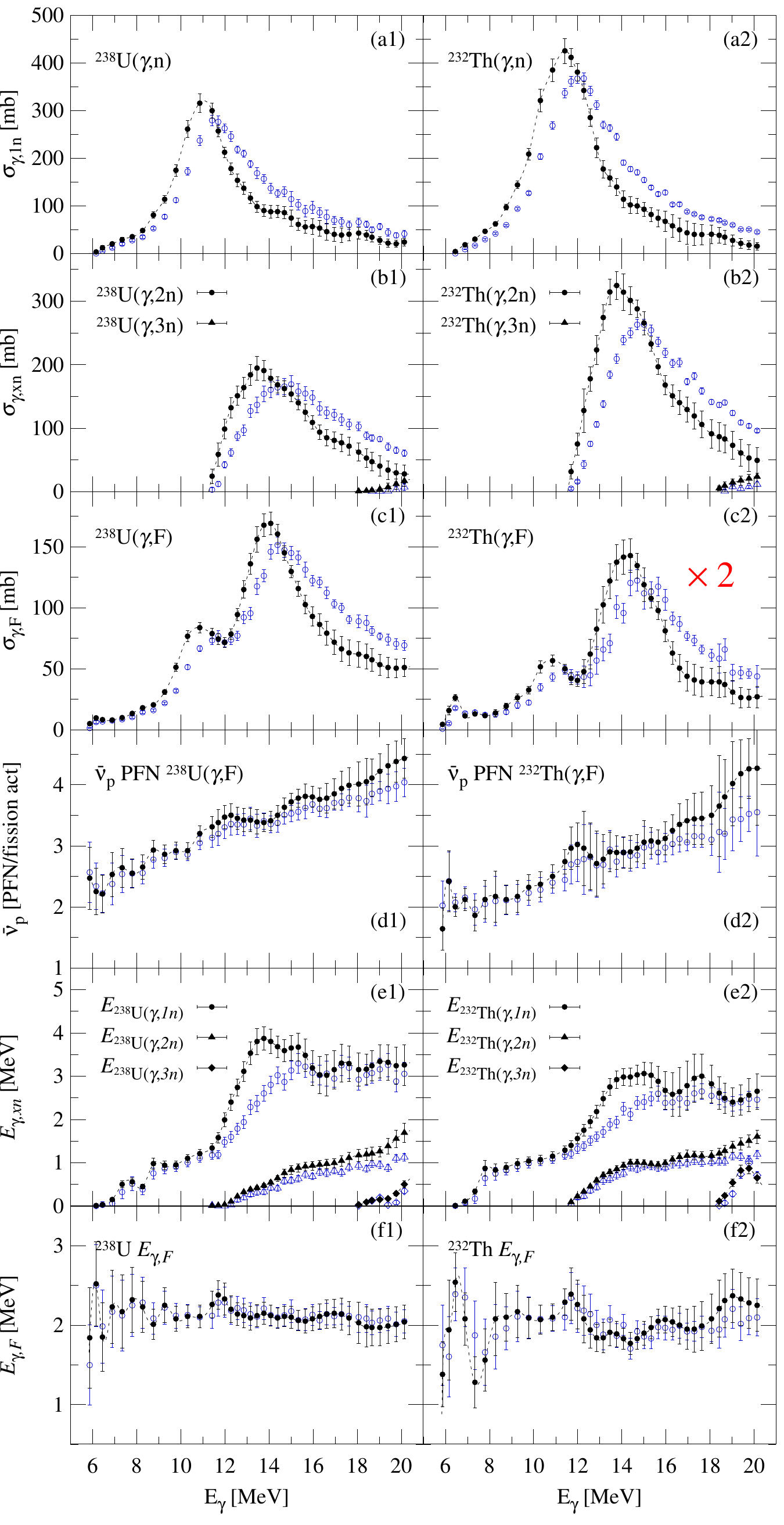} 
\caption{Present (left) $^{238}$U and (right) $^{232}$Th results before (open symbols) and after energy unfolding (full symbols): (a) $(\gamma,\,n)$, (b) $(\gamma,\,2n)$ and $(\gamma,\,3n)$ and (c) $(\gamma,\,F)$ cross sections, (d) mean PFN numbers per fission act, (e) $(\gamma,\,n)$, $(\gamma,\,2n)$ and $(\gamma,\,3n)$ average photoneutron energies and (f) average PFN energies. The error bars account for the statistical uncertainty and for the uncertainties in the neutron detection efficiency, photon flux, target thickness and incident photon spectra.\label{fig_mono_vs_unfolded}}
\end{figure}

The measured quantities discussed above are referred to as monochromatic approximations and represent in fact the folding between the true, energy dependent quantities, and the spectral distribution of the incident photon beams. Explicitly, the measured cross sections are the folding of the excitation function and the beam spectral distribution:
\begin{equation} \label{eq_folded_cs}
\sigma_{\gamma, \,kn}^{\mathrm{fold}}(E_m)= \cfrac{1}{\xi} \int_0^{E_m} L(E_\gamma,E_m) \sigma_{\gamma,kn}(E_\gamma)\,dE_\gamma.
\end{equation}
The measured average energies and mean numbers of PFNs per fission act are the folding between (i) the $E_\gamma$ incident energy dependent functions $E_{\gamma,kn}(E_\gamma)$, $E_{\gamma,F}(E_\gamma)$ and $\overline{\nu}_{p}(E_\gamma)$, (ii) the $L(E_\gamma,\!E_m)$ beam spectral distribution and (iii) the excitation functions of the photoneutron and respectively photofission cross sections:
\begin{align} 
E_{\gamma,kn}^{\mathrm{fold}}(E_m)=& \cfrac{\int_0^{E_m} \!\! E_{\gamma,kn}(E_\gamma) L(E_\gamma,\!E_m)\sigma_{\gamma,kn}(E_\gamma)dE_\gamma}{\sigma_{\gamma,kn}^{\mathrm{fold}}(E_m)\xi} \label{eq_folded_en_gxn} \\ 
E_{\gamma,F}^{\mathrm{fold}}(E_m)= & \cfrac{\int_0^{E_m} \!\! E_{\gamma,F}(E_\gamma) L(E_\gamma,\!E_m) \sigma_{\gamma,F}(E_\gamma)dE_\gamma}{\sigma_{\gamma,F}^{\mathrm{fold}}(E_m)\xi} \label{eq_folded_en_PFN} \\   
\overline{\nu}_{p}^{\mathrm{fold}}(E_m)= & \cfrac{\int_0^{E_m} \!\! \overline{\nu}_{p}(E_\gamma) L(E_\gamma,\!E_m) \sigma_{\gamma,F}(E_\gamma)dE_\gamma}{\sigma_{\gamma,F}^{\mathrm{fold}}(E_m)\xi}. \label{eq_folded_nu}
\end{align}
In the above equations, $L(E_\gamma,E_m)$ distributions are defined as the average path length per unit energy traveled through the target by a $E_\gamma$ photon in an LCS $\gamma$-ray beam of $E_m$ maximum energy. As described in Ref.~\cite{filipescu_NIM_LCS_22}, the $L(E_\gamma,E_m)$ distributions account for the $\gamma$-beam self-attenuation in the target and for the secondary radiation generated by electromagnetic interaction of the $\gamma$-beam with the target, which can have sufficiently high energies to induce nuclear reactions in the target. We have found that, given the limited target areal density and reduced maximum energies of up to 20~MeV employed in the present experiment, the secondary photons had a low contribution to the total spectrum. Nevertheless, it has been taken into account.

Having defined in Eqs.~\eqref{eq_folded_cs}-\eqref{eq_folded_nu} the measured folded quantities, we further apply an iterative energy unfolding procedure described in Refs.~\cite{Renstrom18,LarsenTveten23}. Figure~\ref{fig_mono_vs_unfolded} shows the experimental results in monochromatic approximation, thus before the energy unfolding procedure (empty symbols), and the energy unfolded ones (full symbols) for the (a) $(\gamma,\,n)$, (b) $(\gamma,\,2n)$ and $(\gamma,\,3n)$, (c) $(\gamma,\,F)$ reactions cross sections, (d) mean numbers of PFNs per fission act and the average energies of (e) $(\gamma,\,n)$, $(\gamma,\,2n)$, $(\gamma,\,3n)$ photoneutrons and of (f) PFNs. We note that the energy unfolding procedure hasn't been applied on the PFNs multiplicity distribution width $\sigma$ which shows a slow variation with the increase in excitation energy.

The error bars for the energy unfolded results account for the statistical uncertainties in the neutron detection and for uncertainties of 3$\%$ for the neutron detection efficiency~\cite{Utsunomiya17,Gheorghe2017}, 3$\%$ for the photon flux determination, 0.25$\%$ for the target thickness and the incident photon spectra uncertainty. The uncertainty in the incident photon spectra, shown by the blue bands in Fig.~\ref{fig_inc_spectra_lin2}, has been propagated by applying the unfolding procedure separately for the upper and lower limit of the incident spectra.  

\subsection{Separation of first and second fission chances}

The total photofission cross sections $\sigma_{\gamma,\,F}$ can be expressed in terms of the $\sigma_{\gamma,\,f}$ first- and $\sigma_{\gamma,\,nf}$ second-chance fission components:
\begin{equation}\label{eq_sigma_gF_sum}
\sigma_{\gamma,\,F} = \sigma_{\gamma,\,f} + \sigma_{\gamma,\,nf}.
\end{equation}
In order to describe the relative contributions of both fission chances, we will use the first-chance probability $\texttt{p}$ defined as the ratio of the first-chance photofission cross section and the total photofission cross section:
\begin{equation} \label{eq_p_fis_chance_def}
\texttt{p} = \sigma_{\gamma,\,f} / \sigma_{\gamma,\,F}.
\end{equation}

From Eqs.~\eqref{eq_sigma_gF_sum} and \eqref{eq_p_fis_chance_def} it follows that the $\overline{\nu}_p$ total number of PFNs is given by:
\begin{equation}\label{eq_fis_chance_nu_p}
\overline{\nu}_p = \texttt{p} \overline{\nu}_{\gamma,f} + (1-\texttt{p}) (1 + \overline{\nu}_{\gamma,nf}),
\end{equation}
and the $E_{\gamma,F}$ average energy of the total PFNs spectrum is given by:
\begin{equation}\label{eq_fis_chance_energy}
E_{\gamma,F}\overline{\nu}_p = \texttt{p} E_{\gamma,f} \overline{\nu}_{\gamma,f} + (1-\texttt{p}) [ E_\mathrm{prefiss} + \overline{\nu}_{\gamma,nf} E_{\gamma,nf}],  
\end{equation}
where $E_{\gamma,f}$ and $\overline{\nu}_{\gamma,f}$ are the average energy of PFNs and respectively the mean number of PFNs emitted in the first-chance photofission reactions, $E_{\gamma,nf}$ and $\overline{\nu}_{\gamma,nf}$ are the same but for the second-chance photofission, while $E_\mathrm{prefiss}$ is the average energy of the prefission neutron emitted before the second-chance fission.

We note that Eq.~\eqref{eq_fis_chance_energy} is strictly valid for the neutron emission spectrum. However, in the flat-efficiency approximation in which prefission neutrons of low-energy $E_\mathrm{prefiss}$ are recorded with equal probability as the first- and second-chance PFNs, Eq.~\eqref{eq_fis_chance_energy} is also a good approximation for the recorded neutron spectrum.  

In order to solve Eqs.~\eqref{eq_fis_chance_nu_p} and \eqref{eq_fis_chance_energy} and obtain experimental estimations for the first- and second-chance photofission contributions, we follow the technique employed by the Livermore group~\cite{Caldwell_1980_NSE} and make use of the following assumptions:

\paragraph{} For the first fission chance, there is a linear dependence with the excitation energy for the $E_{\gamma,f}$ average energy of PFNs and for the $\overline{\nu}_{\gamma,f}$ mean number of PFNs emitted per fission act. Thus, the $E_{\gamma,f}$ and $\overline{\nu}_{\gamma,f}$ values at excitation energies above the $B_{nf}$ second-chance fission threshold are obtained by linear extrapolation of their values below the $B_{nf}$.  

\paragraph{} The average energy of the PFNs spectrum is related to the mean number of PFNs as:
\begin{equation}\label{eq_function_PFN_nu_ene}
E_{\gamma,(xn)f} = A_0 + A_1 \cdot (1+\overline{\nu}_{\gamma,(xn)f})^{0.5},
\end{equation}
where $E_{\gamma,(xn)f}$ and $\overline{\nu}_{\gamma,(xn)f}$ are the average energy and mean number of PFNs emitted in the ($x$+1)-chance photofission reaction and which do not include the prefission neutron component. Thus, the $A_0$ and $A_1$ coefficients are determined from the least squares fit to the first-chance photofission $E_{\gamma,f}$ and $\overline{\nu}_{\gamma,f}$ experimental results. Then, the $E_{\gamma,nf}$ average energy of second-chance prompt photofission neutrons from Eq.~\eqref{eq_fis_chance_energy} can be expressed in terms of the $\overline{\nu}_{\gamma,nf}$ mean number of second-chance photofission neutrons:  
\begin{equation}
E_{\gamma,nf} = f(\overline{\nu}_{\gamma,nf}).
\end{equation}

\paragraph{} The $E_\mathrm{prefiss}$ energy of the prefission neutron can be estimated from the experimental $E_{\gamma,\,n}$ average energies of photoneutrons at low excitation energies. 

Following the steps described above for determination of $\overline{\nu}_{\gamma,f}$, $E_{\gamma,f}$ and $E_\mathrm{prefiss}$ at excitation energies above $B_{nf}$ and by expressing the $E_{\gamma,nf}$ as function of $\overline{\nu}_{\gamma,nf}$, the system of Eqs.~\eqref{eq_fis_chance_nu_p} and \eqref{eq_fis_chance_energy} can be numerically solved in order to extract the remaining two variables, which are the $\texttt{p}$ probability of the first-chance photofission reaction and the $\overline{\nu}_{\gamma,nf}$ average number of PFNs emitted in the second-chance photofission reaction. The results are discussed in Section~\ref{sec_res_separation_fis_chance}.

\section{Results and discussion}
\label{sect_results}

\subsection{Prompt fission neutron multiplicities}\label{sec_res_PFN_mult}

\begin{figure}[t]
\includegraphics[width=0.48\textwidth, angle=0]{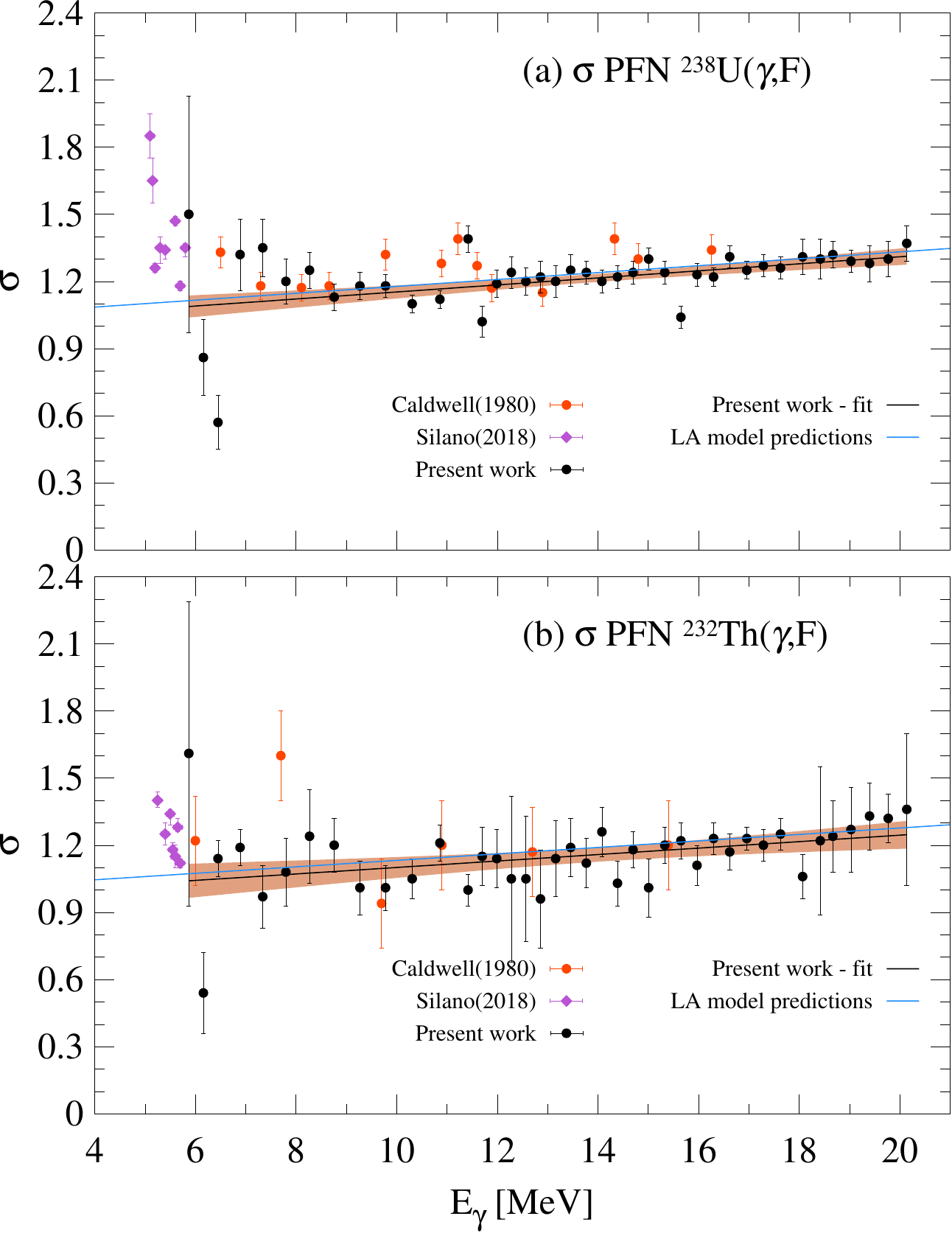} 
\caption{Dependence with incident photon energy for the $\sigma$ spread of PFNs multiplicity distributions in the photofission reactions on (a) $^{238}$U and (b) $^{232}$Th. Present results (full black dots) are compared with LA model predictions~\cite{Tudora_2023_ND} (full blue lines), recent HI$\gamma$S LCS $\gamma$-ray beam data of Silano and Karwowski~\cite{Silano18} (full purple diamonds) and Livermore positron in flight annihilation data~\cite{Caldwell1980_PRC} (full red dots). A linear fit to the present data is shown by the full black line.\label{fig_sigma}} 
\end{figure}

\begin{table}[b]
\caption{\label{table_sigma_fit} Prompt fission neutrons multiplicity width parameter $\sigma$.}
\begin{ruledtabular}
\begin{tabular}{lc}
Target nucleus             & Least squares fit to present data  \\ \colrule
$^{238}\textrm{U}$         & $\sigma = 0.995 + 0.0157 \cdot E_\gamma$      \\   
$^{232}\textrm{Th}$        & $\sigma = 0.956 + 0.0144 \cdot E_\gamma$      \\  
\end{tabular}
\end{ruledtabular}
\end{table}

Figure~\ref{fig_sigma} shows the present (a) $^{238}$U and (b) $^{232}$Th results for the PFNs multiplicity distribution width $\sigma$ compared with all existing data. For both actinide targets, the present values show a slow increase as function of the excitation energy. This is also observed in the results of Caldwell et al. for $^{238}$U, but not for $^{232}$Th, for which a $\sigma$~=~1.183 constant value independent of the excitation energy was obtained. The least squares fit to the present data given in Table~\ref{table_sigma_fit} and plotted as a black line in Fig.~\ref{fig_sigma} is well reproduced by LA model predictions~\cite{Tudora_2023_ND} shown by the blue line in Fig.~\ref{fig_sigma}. The recent low-energy data of Silano and Karwowski however do not confirm the energy increase in $\sigma$.

\begin{figure}[t]
\includegraphics[width=0.48\textwidth, angle=0]{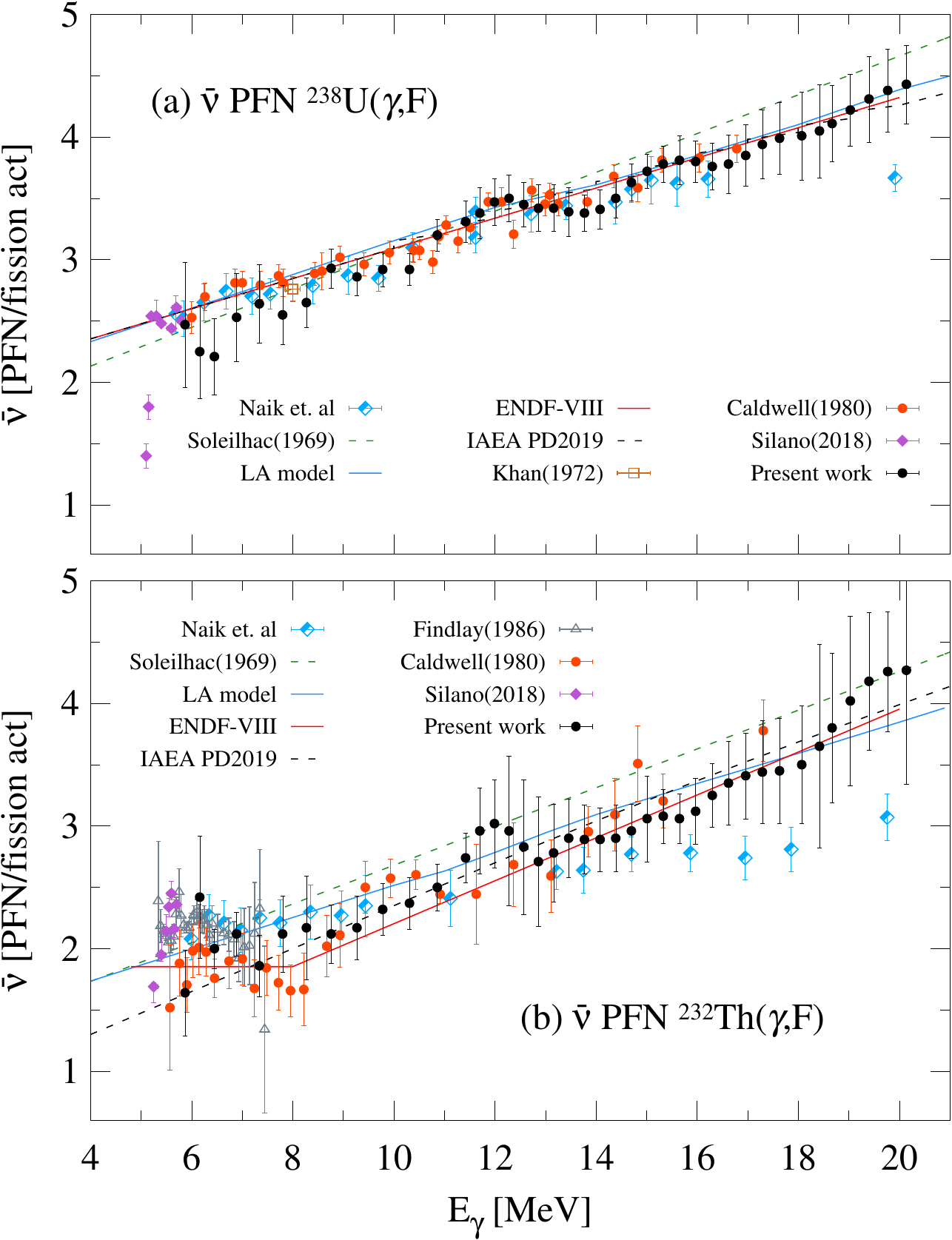} 
\caption{Dependence with incident photon energy for the mean PFN multiplicities in the photofission reactions on (a) $^{238}$U and (b) $^{232}$Th. The present results (full black dots) are compared with recent HI$\gamma$S LCS $\gamma$-ray beam data of Silano and Karwowski~\cite{Silano18} (full purple diamonds), Livermore positron in flight annihilation data~\cite{Caldwell1980_PRC} (full red dots), capture $\gamma$-ray data~\cite{Khan1972} (brown empty squares), bremsstrahlung data of Ref.~\cite{Findlay1986} (gray open triangles) and indirect determinations from bremsstrahlung studies of fission product yield distributions of Refs.~\cite{Chattopadhyay1973,Piessens1993,Naik2011,Naik2012a,Naik2012b,Naik2015} (half full blue diamonds). The green dashed-dotted lines show the systematic linear dependences deduced from neutron-induced fission experiments by Soleilhac et al.~\cite{Soleilhac1969} and used by the Saclay group~\cite{Veyssiere1973} in the data reduction. The full blue lines are LA model predictions~\cite{Tudora_2023_ND}. The IAEA PD 2019~\cite{Kawano2020} and the ENDF/B-VIII.0~\cite{Brown2018} evaluations are shown in dotted black and respectively solid red lines. \label{fig_avgPFN}}
\end{figure}

\begin{figure*}[t]
\includegraphics[width=0.98\textwidth, angle=0]{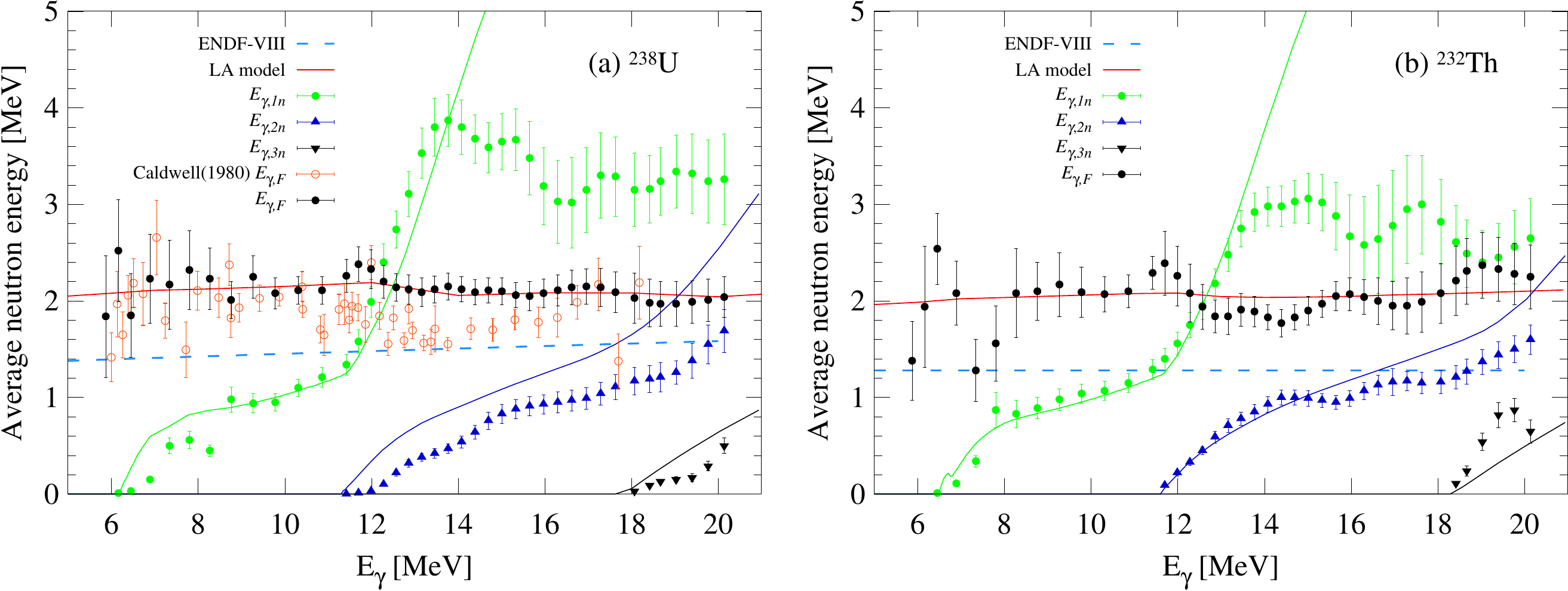} 
\caption{Average energies of neutrons emitted in photon induced reactions on (a) $^{238}$U and (b) $^{232}$Th. $E_{\gamma,\,n}$ average energy of neutrons emitted in $(\gamma,\,n)$ reactions (green dots), $E_{\gamma,\,2n}$ for $(\gamma,\,2n)$ neutrons (blue triangles) and $E_{\gamma,\,3n}$ for $(\gamma,\,3n)$ neutrons (black triangles) are compared with results of present EMPIRE statistical model calculations (solid lines in corresponding color for each reaction). Present $E_{\gamma,\,F}$ PFNs energies (black dots) are compared to the ENDF/B-VIII.0 evaluation~\cite{Brown2018} (dashed blue lines) and LA model predictions~\cite{Tudora_2023_ND} (red lines), and, for $^{238}$U, also to the results of Caldwell et al.~\cite{Caldwell1980_PRC} (empty red dots).\label{fig_energy_unfolded}}
\end{figure*}

Figure~\ref{fig_avgPFN} shows the present experimental mean numbers of PFNs per fission act for photofission reactions on (a) $^{238}$U and (b) $^{232}$Th compared with existing data, evaluations and predictions. The present data confirm the previous general observations~\cite{Caldwell1980_PRC,Naik2015} that the $\overline{\nu}_p$ increases with increasing incident energy and that, at the same excitation energy, the $\overline{\nu}_p$ values are higher for $^{238}$U than for $^{232}$Th because of the increase in the mass and charge of the fissioning system. 

For $^{238}$U, there is a generally good agreement between all data sets. For excitation energies below 9~MeV, the present $\overline{\nu}_p$ are slightly lower than the previous results of Caldwell et al.~\cite{Caldwell1980_PRC}, which connect well to the sub-barrier data of Silano and Karwowski~\cite{Silano18}. However, the present data are in good consistency with the Livermore ones in the higher excitation energy region. The experimental results are well reproduced by the IAEA PD 2019~\cite{Kawano2020} and the ENDF/B-VIII.0~\cite{Brown2018} evaluations and by the LA model predictions~\cite{Tudora_2023_ND}. The predictions obtained from the neutron induced fission experiments of Soleilhac et al.~\cite{Soleilhac1969} shown an energy increase steeper than the experimental data. 

For $^{232}$Th, a more complex relationship between the $\overline{\nu}_p$ and the excitation energy is observed in both the present data and the previous results of Caldwell et al.~\cite{Caldwell1980_PRC}. Although there is good agreement between the overall increase slope and magnitude of the present and the Livermore results, the two data sets do not agree in describing the individual resonant structures observed in the $\overline{\nu}_p$ energy dependency. We notice that the $\overline{\nu}_p$ error bars are larger for $^{232}$Th than for $^{238}$U in both the present and the Livermore data sets. The increased uncertainty in the determination of fission related parameters for $^{232}$Th originates from the poor statistics owing to the small size of $\sigma_{\gamma,\,F}$ in $^{232}$Th, which follows from the low fissility parameter $\sim{Z^2/A}$. Considering the large error bars of the experimental data, one can consider that the IAEA PD 2019~\cite{Kawano2020} and the ENDF/B-VIII.0~\cite{Brown2018} evaluations and also the LA model predictions~\cite{Tudora_2023_ND} reproduce well the experimental $\overline{\nu}_p$ values in the $^{232}$Th$(\gamma\,F)$ reaction. However, the Soleilhac et al.~\cite{Soleilhac1969} systematic which was used in the neutron multiplicity sorting procedure applied by the Saclay group~\cite{Veyssiere1973} overestimates both the present and the Livermore~\cite{Caldwell1980_PRC} experimental data for excitation energies above $\sim$7~MeV.  

\begin{figure*}[t]
\includegraphics[width=0.98\textwidth, angle=0]{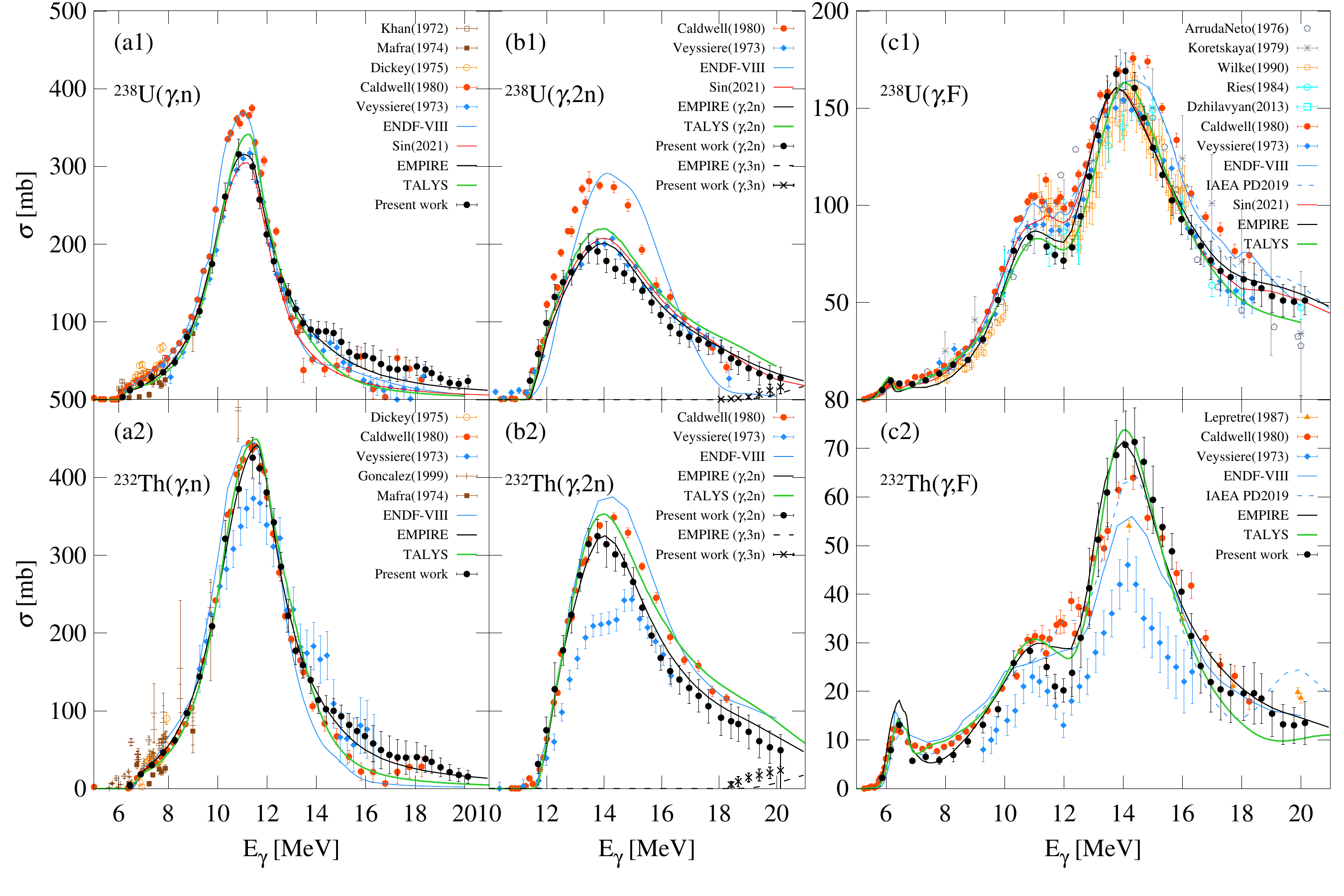} 
\caption{Present cross sections (black dots) for the (a) $(\gamma,\,n)$, (b) $(\gamma,\,2n)$ and $(\gamma,\,3n)$ and (c) $(\gamma,\,F)$ reactions on (upper row) $^{238}$U and (lower row) $^{232}$Th compared with existing data obtained with positron in flight annihilation beams at Saclay~\cite{Veyssiere1973} (blue full diamonds), Livermore~\cite{Caldwell1980_PRC} (red full dots), Giessen~\cite{Ries1984} (cyan empty dots) and Moscow~\cite{Dzhilavyan2013} (cyan empty square), bremsstrahlung beams~\cite{ArrudaNeto1976,Koretskaya1979} (gray symbols), bremsstrahlung monochromators~\cite{Dickey1975,Wilke1990,Lepretre1987} (orange symbols), capture $\gamma$-rays~\cite{Khan1972,Mafra1974,Goncalez1999} (brown symbols). Present EMPIRE and TALYS statistical model calculations and the $^{238}$U ones of Ref.~\cite{msin_2021} are shown by the black, green and red solid lines, respectively. \label{fig_cs_partiale}}
\end{figure*}

\subsection{Average energies of photoneutrons and PFNs}\label{sec_average_energy_PFN}

Figure~\ref{fig_energy_unfolded} shows the average energies of neutrons emitted in photoneutron $(\gamma,\,1-3n)$ reactions and of PFNs emitted in photofission reactions on (a) $^{238}$U and (b) $^{232}$Th. For both nuclei, the average energy of $(\gamma,\,n)$ neutrons shows a slow rise above the $S_n$ followed by a sharp increase above the $S_{2n}$ and then a slow decrease above incident energies of 14~MeV. The EMPIRE calculations describe well the increasing $E_{\gamma,\,n}$ behavior but do not follow the slow decrease above 14~MeV excitation energy. The average energy of $(\gamma,\,2n)$ neutrons rises slowly above the $S_{2n}$ up to stable value of $\sim$1~MeV and then continues to increase at excitation energies above $S_{3n}$. Both the $E_{\gamma,\,2n}$ and $E_{\gamma,\,3n}$ experimental values are reasonably well reproduced by the EMPIRE calculations.

The $E_{\gamma,\,F}$ average energy of the PFNs shows a slow rise with increase excitation energies above the fission barrier. A drop in the average energy at $S_{2n}$ is observed for both nuclei, however more pronounced for $^{232}$Th. The $E_{\gamma,\,F}$ continues to rise slowly above the $S_{2n}$. The experimental data are well reproduced by the LA model predictions~\cite{Tudora_2023_ND} (red lines). The ENDF/B-VIII.0 evaluation~\cite{Brown2018} (blue lines) strongly underestimates the experimental $E_{\gamma,\,F}$ values. 

\begin{figure*}[t]
\includegraphics[width=0.8\textwidth, angle=0]{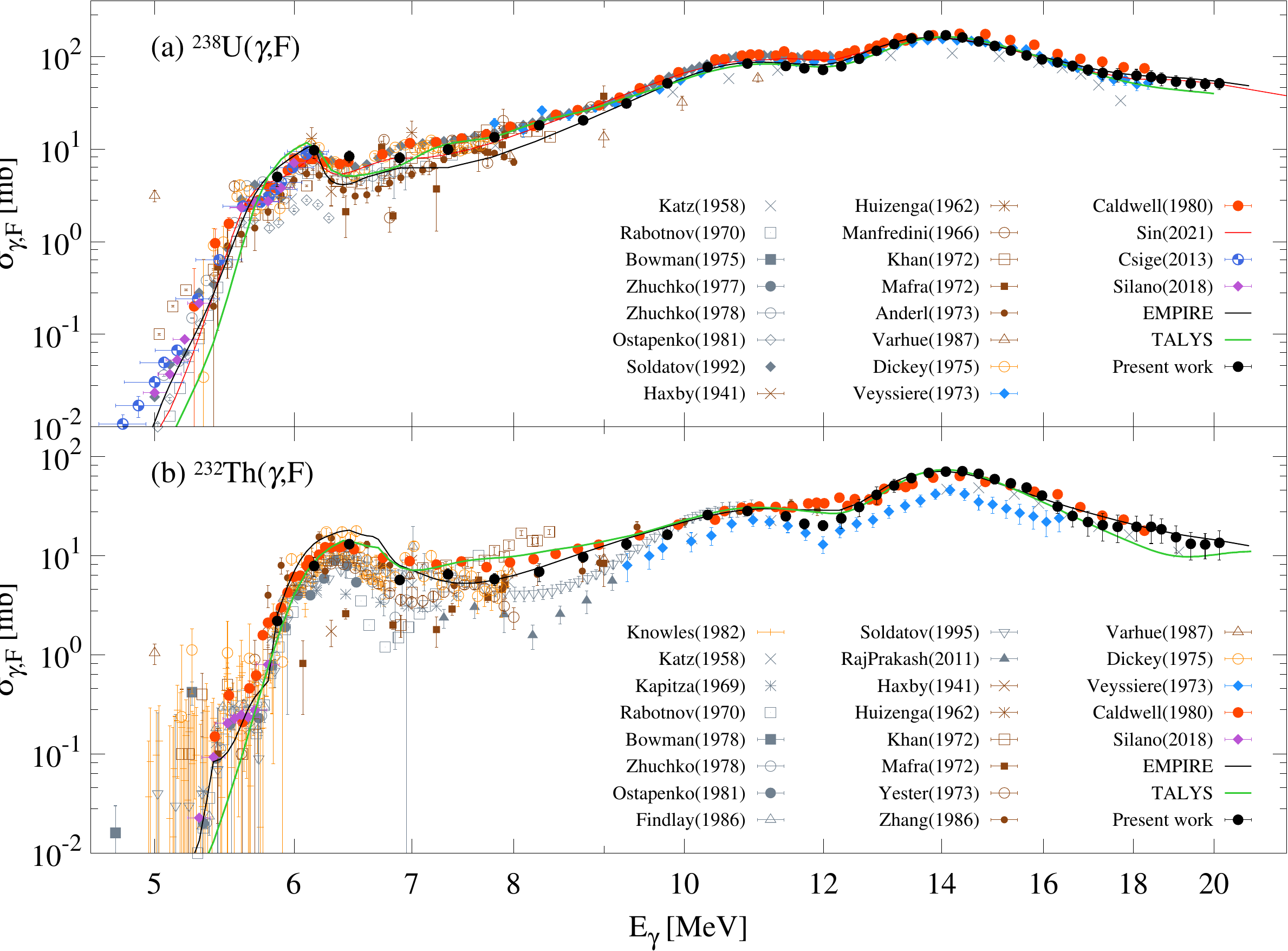} 
\caption{Present photofission cross sections (black full dots) for (a) $^{238}$U and (b) $^{232}$Th compared in the low energy region to present EMPIRE and TALYS statistical model calculations (black and green lines, respectively), the $^{238}$U ones of Ref.~\cite{msin_2021} (red line), and to existing data: recent LCS $\gamma$-ray beam data obtained at HI$\gamma$S by Csige et al.~\cite{Csige2013} (blue half empty dots) using a fission chamber and by Silano and Karwowski~\cite{Silano18} (full purple diamonds) using a moderated $^3$He detection array, the Saclay~\cite{Veyssiere1973} (blue full diamonds) and Livermore~\cite{Caldwell1980_PRC} (red full dots) positron in flight annihilation data, bremsstrahlung data~\cite{Katz1958,Kapitza1969,Rabotnov1970,Bowman1975,Bowman1978,Zhuchko1977,Zhuchko1978,Ostapenko1981,Findlay1986,Soldatov1992,Soldatov1995,RajPrakash2011} (gray symbols), bremsstrahlung monocromator data~\cite{Dickey1975,knowles_1982} (orange symbols), capture $\gamma$-ray data~\cite{Haxby1941,Huizenga1962,Manfredini1966,Khan1972,Mafra1972,Yester1973,Anderl1973,Zhang1986,Varhue1987} (brown symbols).\label{fig_cs_gF_low}}
\end{figure*}

\begin{figure*}[t]
\includegraphics[width=0.98\textwidth, angle=0]{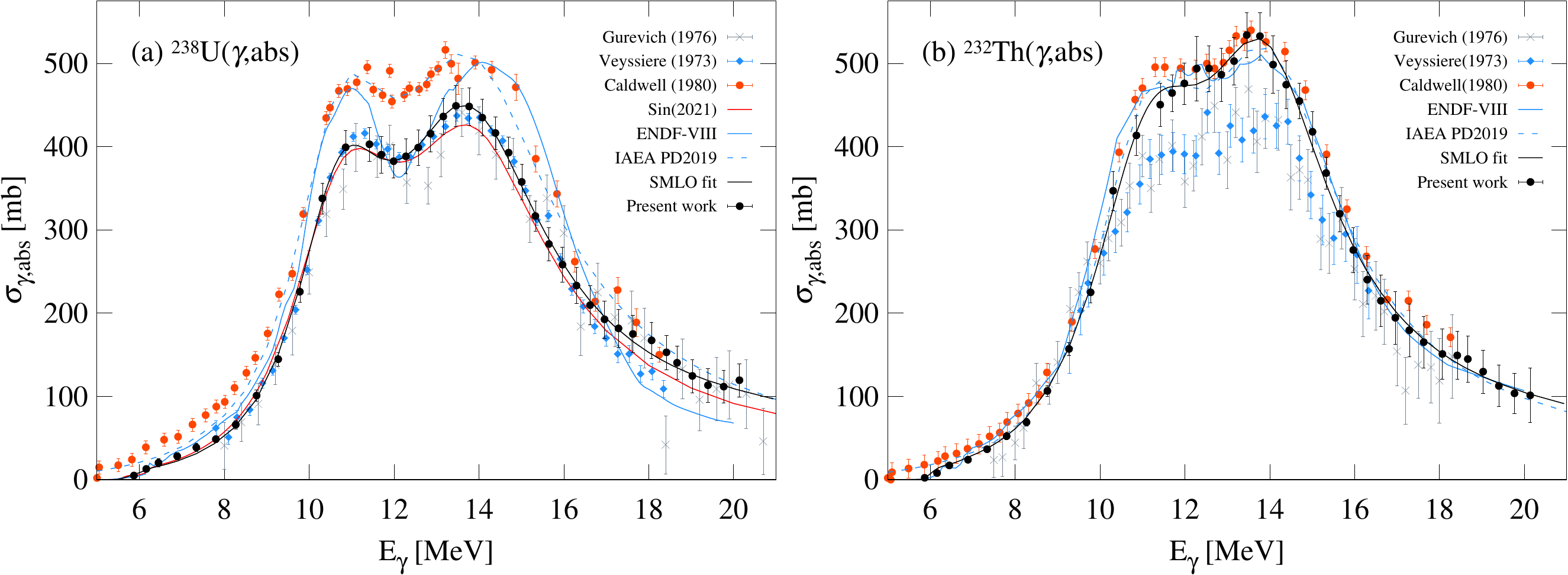} 
\caption{Present photoabsorption cross sections for (a) $^{238}$U and (b) $^{232}$Th compared to existing data obtained with positron in flight annihilation beams at Saclay~\cite{Veyssiere1973} (blue full diamonds) and at Livermore~\cite{Caldwell1980_PRC} (red full dots) and to bremsstrahlung beam data~\cite{Gurevich1976} (gray). The SMLO curves correspond to fits to the present data using the Simple Modified Lorentzian function described in Ref.~\cite{plujko_2018}. We also show the Sin et al.~\cite{msin_2021} photoabsorption cross section (red line) and the ENDF-VIII (solid blue lines) and PD2019 (dotted blue lines) evaluations.\label{fig_cs_abs}}
\end{figure*}

\subsection{Photoneutron and photofission cross sections}

Present cross section results for $^{238}$U~(top) and $^{232}$Th~(bottom) are shown in Fig.~\ref{fig_cs_partiale} in comparison with existing data, the ENDF/B-VIII.0 evaluation~\cite{Brown2018} as well as statistical calculations. All available literature data sets are plotted for the $\sigma_{\gamma,\,n}$ and $\sigma_{\gamma,\,2n}$ photoneutron cross sections, while the for the photofission reaction we have plotted only the data sets which extend in the GDR region. 

\subsubsection{$(\gamma,\,in)$ and $(\gamma,\,F)$ reactions in $^{238}$U}

The present photoneutron $\sigma_{\gamma,\,n}$ cross section results for $^{238}$U plotted in Fig.~\ref{fig_cs_partiale}(a1) show good agreement in the peak energy region with the Saclay data of Veyssiere et al.~\cite{Veyssiere1973} and underestimate by $\sim$20$\%$ the Livermore data of Caldwell et al.~\cite{Caldwell1980_PRC}. At high excitation energies above $\sim$14~MeV, the present $\sigma_{\gamma,\,n}$ cross sections show a rather slow decrease and above 17~MeV stabilize at $\sim$50~mb, which is higher than both the Livermore and Saclay results which show a steeper slope and stabilize close to zero.  

The present $\sigma_{\gamma,\,2n}$ cross sections for $^{238}$U shown in Fig.~\ref{fig_cs_partiale}(b1) are also below the Livermore results of Caldwell et al.~\cite{Caldwell1980_PRC} and in generally good agreement with the Saclay results, especially on the rising slope just above $S_{2n}$, while in the peak energy region and above we obtain slightly lower values. At $\sim$20~MeV excitation energy, above the $S_{3n}$, the present $\sigma_{\gamma,\,2n}$ stabilizes at $\sim$33~mb. 

Figure~\ref{fig_cs_partiale}(c1) shows an overall good agreement between the present and the existing experimental photofission cross sections in $^{238}$U. In consistency with the photoneutron channels, the present $^{238}$U$(\gamma,\,F)$ reaction cross sections are also in better agreement with the Saclay results than with Livermore ones. However, we obtained lower and respectively higher cross sections for the first and respectively second peak. A more pronounced peak separation is observed in the present data compared with the existing positron in flight annihilation ones, which is also present in the bremsstrahlung monochromator data of Wilke and coworkers~\cite{Wilke1990}. At high excitation energies of $\sim$20~MeV, the present cross sections stabilize to $\sim$50~mb. 

\subsubsection{$(\gamma,\,in)$ and $(\gamma,\,F)$ reactions in $^{232}$Th}

The present $\sigma_{\gamma,\,n}$ cross sections for $^{232}$Th plotted in Fig.~\ref{fig_cs_partiale}(a2) are in overall good agreement with both the Saclay and Livermore cross sections. In the low energy region, they are below the bremsstrahlung monochromator data of Dickey et al.~\cite{Dickey1975} and above the capture $\gamma$-ray data of Mafra et al.~\cite{Mafra1974}, situation observed also in the $^{238}$U$(\gamma,\,n)$ cross sections. Above 14~MeV excitation energy, the present cross section shows a slow decrease and stabilizes at $\sim$35~mb. 

The present $^{232}$Th$(\gamma,\,2n)$ cross sections plotted in Fig.~\ref{fig_cs_partiale}(b2) are in good agreement with the Livermore data on the rising slope above $S_{2n}$. For excitation energies above 14~MeV, present $\sigma_{\gamma,\,2n}$ are systematically lower than the Livermore cross sections and show a continuous decrease without stabilizing on a constant plateau. The present results do not confirm the sharp increase observed at $\sim$15~MeV in the Saclay data. 

Present $^{232}$Th$(\gamma,\,F)$ cross sections shown in Fig.~\ref{fig_cs_partiale}(c2) are generally higher than the Saclay results of Veyssiere et al.~\cite{Veyssiere1973}. The present data are in good agreement with the Livermore results of Caldwell et al.~\cite{Caldwell1980_PRC}, except for the high energy peak region of 13~to~15~MeV excitation energy where they are higher than the Livermore data. The present $\sigma_{\gamma,\,F}$ show a steep drop from the high energy peak and stabilize to $\sim$17~-~18~mb at excitation energies above 17~MeV. As observed also in the case of $\overline{\nu}_p$, the relative uncertainties of the $^{232}$Th$(\gamma,\,F)$ cross sections are higher than the $^{238}$U ones, because of the low counting statistics associated to the low photofission cross sections in $^{232}$Th. 

We notice that the IAEA PD 2019~\cite{Kawano2020} and the ENDF/B-VIII.0~\cite{Brown2018} evaluations reproduced the Livermore photonuclear reaction cross sections on both $^{238}$U and $^{232}$Th actinides. Thus, they overestimate the present $^{238}$U photoneutron results while describing reasonably well the photofission channel. For $^{232}$Th, the two recommendations overestimate the $(\gamma,\,2n)$ cross section and underestimate the photofission one. 

\subsubsection{Photofission cross section in the vicinity of $S_n$}

Figure~\ref{fig_cs_gF_low} shows the photofission cross sections on (a) $^{238}$U and (b) $^{232}$Th in log-log scale for good visualization of the region spanning from sub-barrier and up to several MeVs above the $S_n$. A wealth of experimental data is present in this low energy region, obtained with bremsstrahlung~\cite{Katz1958,Kapitza1969,Rabotnov1970,Bowman1975,Bowman1978,Zhuchko1977,Zhuchko1978,Ostapenko1981,Findlay1986,Soldatov1992,Soldatov1995,RajPrakash2011}, bremsstrahlung monochromator~\cite{Dickey1975,knowles_1982}, capture $\gamma$-rays~\cite{Haxby1941,Huizenga1962,Manfredini1966,Khan1972,Mafra1972,Yester1973,Anderl1973,Zhang1986,Varhue1987}. The positron in flight annihilation data of Caldwell et al.~\cite{Caldwell1980_PRC} also extend to energies as low as 5.3~MeV. Two investigations with monochromatic LCS $\gamma$-ray beams have been performed at the HI$\gamma$S facility by Csige et al.~\cite{Csige2013} on $^{238}$U using a stack of 40 thin metallic actinide targets and a fission chamber and, more recently, by Silano and Karwowski~\cite{Silano18} on both $^{238}$U and $^{232}$Th using a moderated array of $^3$He counters. 

The present results are in overall good agreement with the existing data. We reproduce the hump at 6~MeV in $^{238}$U and at 6.5~MeV in $^{232}$Th. On the left side of the 6~MeV hump, there is good overlap between the present results on $^{238}$U photofission cross sections and the LCS $\gamma$-ray beam ones of Csige et al.~\cite{Csige2013} and of Silano and Karwowski~\cite{Silano18}. For $^{232}$Th, the two lowest energy points of the present data connect nicely with the highest energy points of the Silano and Karwowski~\cite{Silano18} data set. 
 
\subsubsection{Total photoabsorption cross sections}

Figure~\ref{fig_cs_abs} shows the photoabsorption cross sections for (a) $^{238}$U and (b) $^{232}$Th. The present photoabsorption cross sections are obtained as the sum of the partial photoneutron and photofission cross sections. For $^{238}$U, they are in good agreement with the Saclay results of Veyssiere et al.~\cite{Veyssiere1973} and with the bremsstrahlung data of Gurevich et al.~\cite{Gurevich1976}. The IAEA PD 2019~\cite{Kawano2020} evaluation reproduced the Livermore results of Caldwell et al.~\cite{Caldwell1980_PRC}, and thus overestimate the present results. The statistical model calculations of Sin et al.~\cite{msin_2021} reproduce well the present data in the low energy and GDR peak regions, but however underestimate the cross sections in the high energy region above 17~MeV. The present $^{232}$Th$(\gamma,\,abs)$ cross sections are in good agreement with the ones of Caldwell et al.~\cite{Caldwell1980_PRC} and thus well reproduced by the IAEA PD 2019~\cite{Kawano2020} and ENDF/B-VIII.0~\cite{Brown2018} evaluations.

\subsection{Separation of fission chances}\label{sec_res_separation_fis_chance}

\begin{figure}[t]
\includegraphics[width=0.49\textwidth, angle=0]{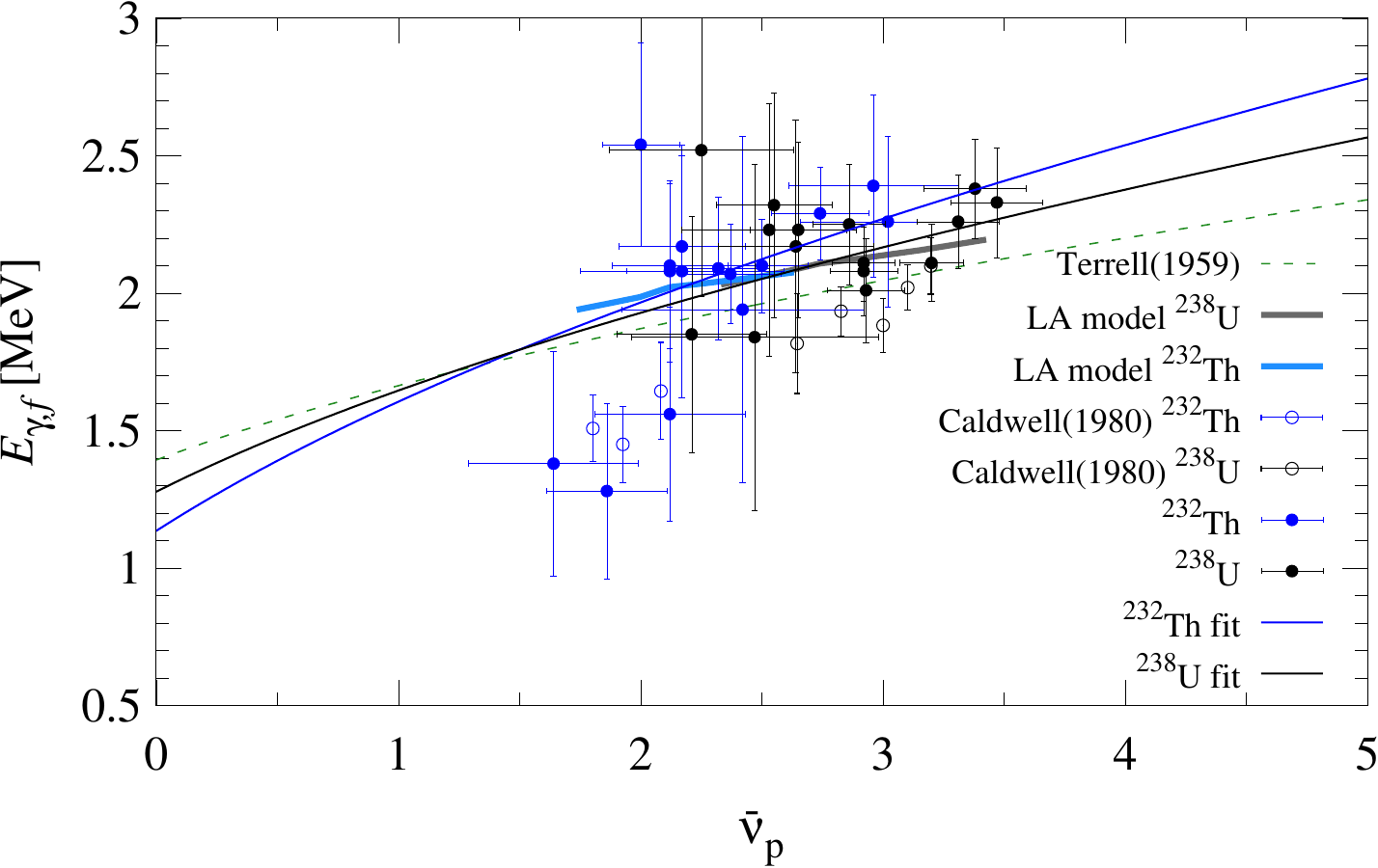} 
\caption{Average energy of PFNs emitted in first-chance photofission reactions on $^{238}$U (black) and $^{232}$Th (blue) vs the $\overline{\nu}_p$ mean number of PFNs. Present experimental results are shown by full dots and their least square fit by solid lines. LA model predictions~\cite{Tudora_2023_ND} are given in dotted lines and the general approximation given by J.~Terrell~\cite{Terrell_1959} is shown by the green dashed-dotted line. \label{fig_enn_f_nu}}
\end{figure}

The average energies of PFNs emitted in photofission reactions at excitation energies below $B_{nf}$ are shown in Fig.~\ref{fig_enn_f_nu}, plotted as a function of $\overline{\nu}_p$. The $\overline{\nu}_p$ and $E_{\gamma,f}$ values have been discussed in Sections~\ref{sec_res_PFN_mult}~and~\ref{sec_average_energy_PFN} and have been shown as functions of the excitation energy in Figs.~\ref{fig_avgPFN}~and~\ref{fig_energy_unfolded}. In Fig.~\ref{fig_enn_f_nu}, the results for each of the actinide targets are shown with separate colors: black for $^{238}$U and blue for $^{232}$Th. A range in $\overline{\nu}_p$ from 1.6 to 3.0 PFNs per fission act is covered for $^{232}$Th and from 2.2 to 3.5 for $^{238}$U. The thin solid lines are least square fits to the experimental data for each nucleus performed using the function given in Eq.~\eqref{eq_function_PFN_nu_ene}. The thick solid lines represent LA model predictions~\cite{Tudora_2023_ND} for each fissioning system. The green dotted line is the evaporation-model prediction of Terrell~\cite{Terrell_1959}. The empty dots are the experimental results of Caldwell et al. retrieved from Fig.~1 of Ref.~\cite{Caldwell1980_PRC}. For $^{232}$Th, we obtain a faster rise in the average PFNs energy with the mean multiplicity when compared to the $^{238}$U results and to the prediction of Terrell, which is in agreement with the Livermore results.

\begin{figure}[t]
\includegraphics[width=0.48\textwidth, angle=0]{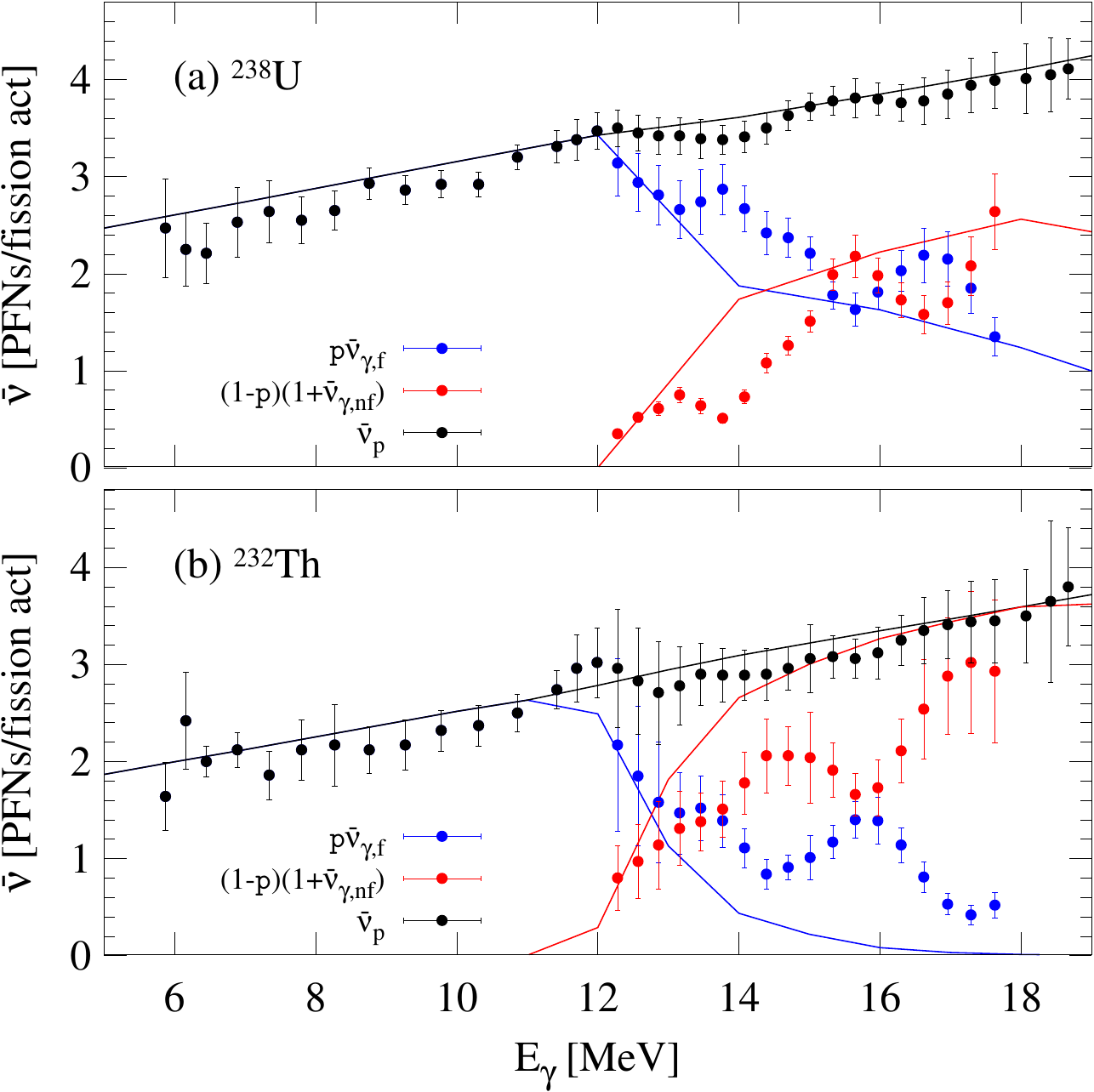} 
\caption{Dependence with incident photon energy for the mean PFNs multiplicities in the photofission reactions on (a) $^{238}$U and (b) $^{232}$Th. The present results for the $\overline{\nu}_p$ total mean number of PFNs per fission act (full black dots) and for the second chance contribution including the prefission neutron (red full dots) are compared with the corresponding LA model predictions~\cite{Tudora_2023_ND}. \label{fig_avgPFN_fis_chance}}
\end{figure}

The $\overline{\nu}_p$ total mean number of PFNs per fission act are compared in Fig.~\ref{fig_avgPFN_fis_chance} with the contribution of the first- and second-chance fission for (a) $^{238}$U and (b) $^{232}$Th. The experimental results (full dots) are compared with LA model predictions~\cite{Tudora_2023_ND} weighted with the present EMPIRE calculations for fission chance probabilities.   

\begin{figure}[t]
\includegraphics[width=0.48\textwidth, angle=0]{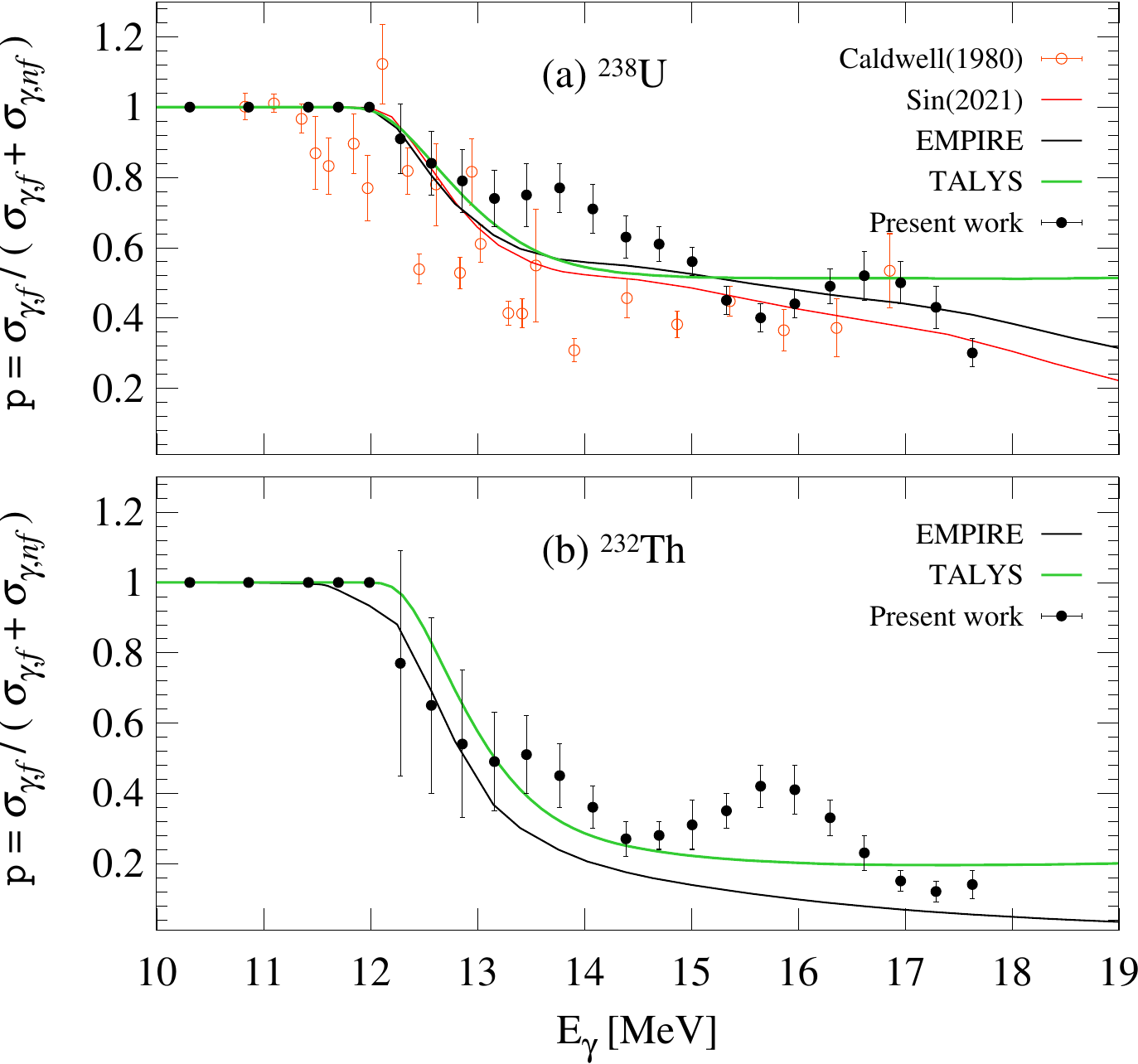} 
\caption{Ratios of the first-chance photofission cross section to the total photofission cross section 
given by the sum of $\sigma_{\gamma,\,f}$ first- and $\sigma_{\gamma,\,nf}$ second- chance photofission cross sections for (a) $^{238}$U and (b) $^{232}$Th. Present data (full black dots) are compared with the results of Caldwell et al.~\cite{Caldwell1980_PRC} (red empty dots) and to present EMPIRE (black lines), TALYS (green lines), and Sin et al.~\cite{msin_2021} (red line) statistical model calculations. \label{fig_p_fis_chance}}
\end{figure}

The $\texttt{p}$ ratio of the first-chance photofission cross section $\sigma_{\gamma,f}$ to the total photofission cross section $\sigma_{\gamma,F}=\sigma_{\gamma,f}+\sigma_{\gamma,nf}$ is shown in Figs.~\ref{fig_p_fis_chance}(a) and~\ref{fig_p_fis_chance}(b) for $^{238}$U and $^{232}$Th, respectively. The present results for $^{238}$U are compared with the data of Caldwell et al.~\cite{Caldwell_1980_NSE,Caldwell1980_PRC}. The present $^{238}$U data confirm the 12.3~MeV value for the $B_{nf}$, while the Caldwell data indicated a lower value for the second-chance fission barrier in $^{238}$U. The statistical model calculations shown by the solid lines in Fig.~\ref{fig_p_fis_chance} reproduce well the average decrease slope of $\texttt{p}$ in both nuclei, but cannot describe the structures at 13.5 and 17~MeV in $^{238}$U and the one at 15.8~MeV in $^{232}$Th. We notice that the probability for the first-chance photofission reaction in $^{238}$U shows a slow decrease with the increase in the excitation energy and retains significant values above $\sim$40$\%$ on the entire $B_{nf}$~--~$S_{3n}$ energy region. On the other hand, the present $^{232}$Th indicate a $B_{nf}$ value slightly lower than the 12.6~MeV one listed in Table~\ref{table_isotopes}. Also, the first-chance photofission probability in $^{232}$Th falls sharply above $B_{nf}$ and reaches $\sim$10$\%$ at excitation energies close to $S_{3n}$. 

\begin{figure}[t]
\includegraphics[width=0.48\textwidth, angle=0]{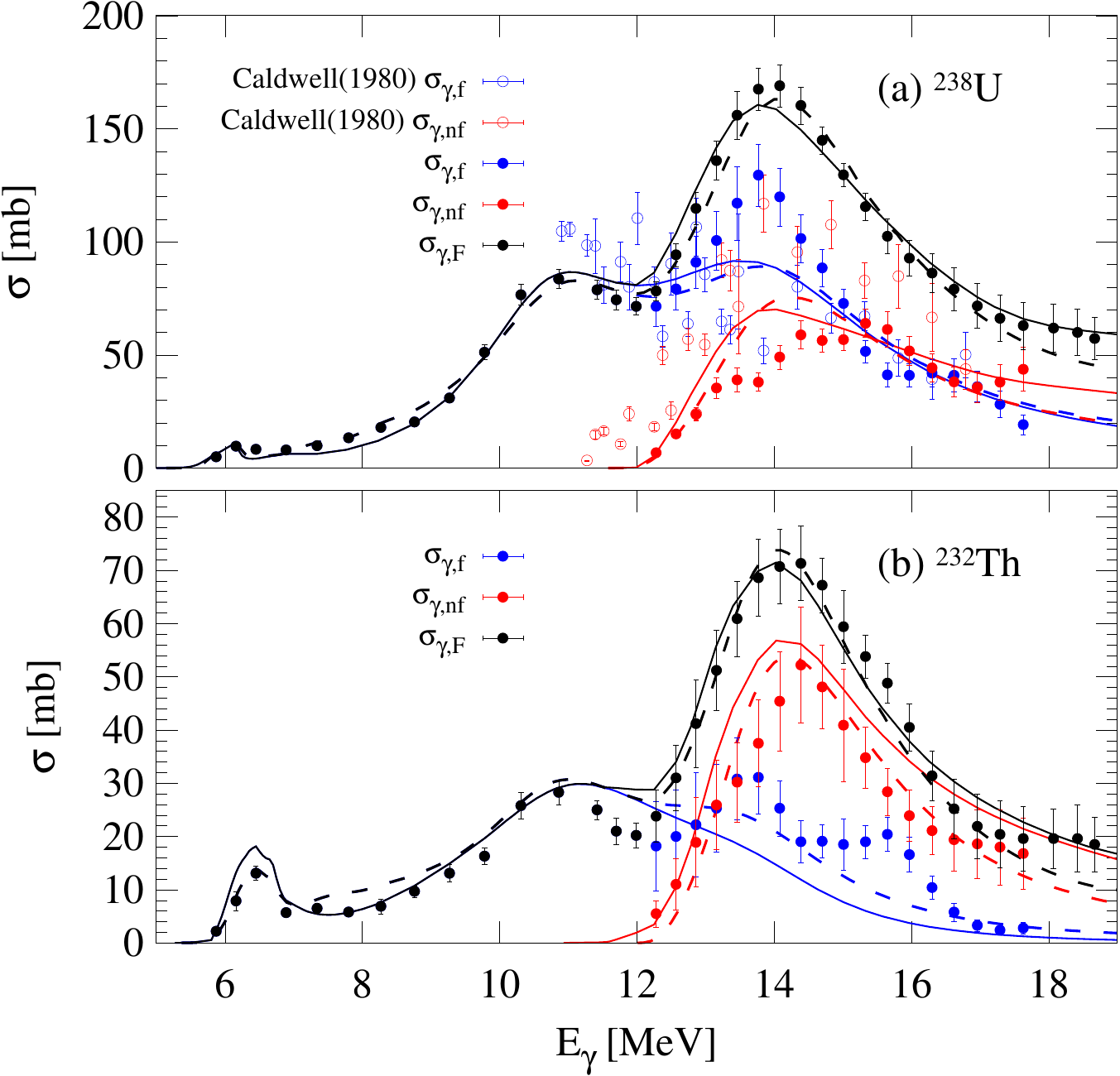} 
\caption{Photofission cross sections for (a) $^{238}$U and (b) $^{232}$Th: total photofission cross sections $\sigma_{\gamma,F}$ (black), first-chance (blue) and second-chance (red) photofission cross sections. Present measurements (full dots) are compared to experimental results of Caldwell et al.~\cite{Caldwell1980_PRC} (empty dots) and present EMPIRE and TALYS statistical model calculations (solid and dashed lines, respectively). \label{fig_csF_fis_chance}}
\end{figure}

Figure~\ref{fig_csF_fis_chance} shows the first-chance (blue) and second-chance (red) photofission cross sections for (a) $^{238}$U and (b) $^{232}$Th as obtained from the total photofission cross section plotted in Fig.~\ref{fig_cs_partiale}(c1,2) and again in Fig.~\ref{fig_csF_fis_chance} (black) and from the first-chance fission probabilities $\texttt{p}$. We note that the pronounced second hump observed in the present first-chance photofission cross section on $^{238}$U is not confirmed by the existing data of Caldwell et al.~\cite{Caldwell1980_PRC} and not reproduced by the present EMPIRE and TALYS statistical model calculations nor by the ones of Sin et al.~\cite{msin_2021}. 

\begin{figure}[t]
\includegraphics[width=0.47\textwidth, angle=0]{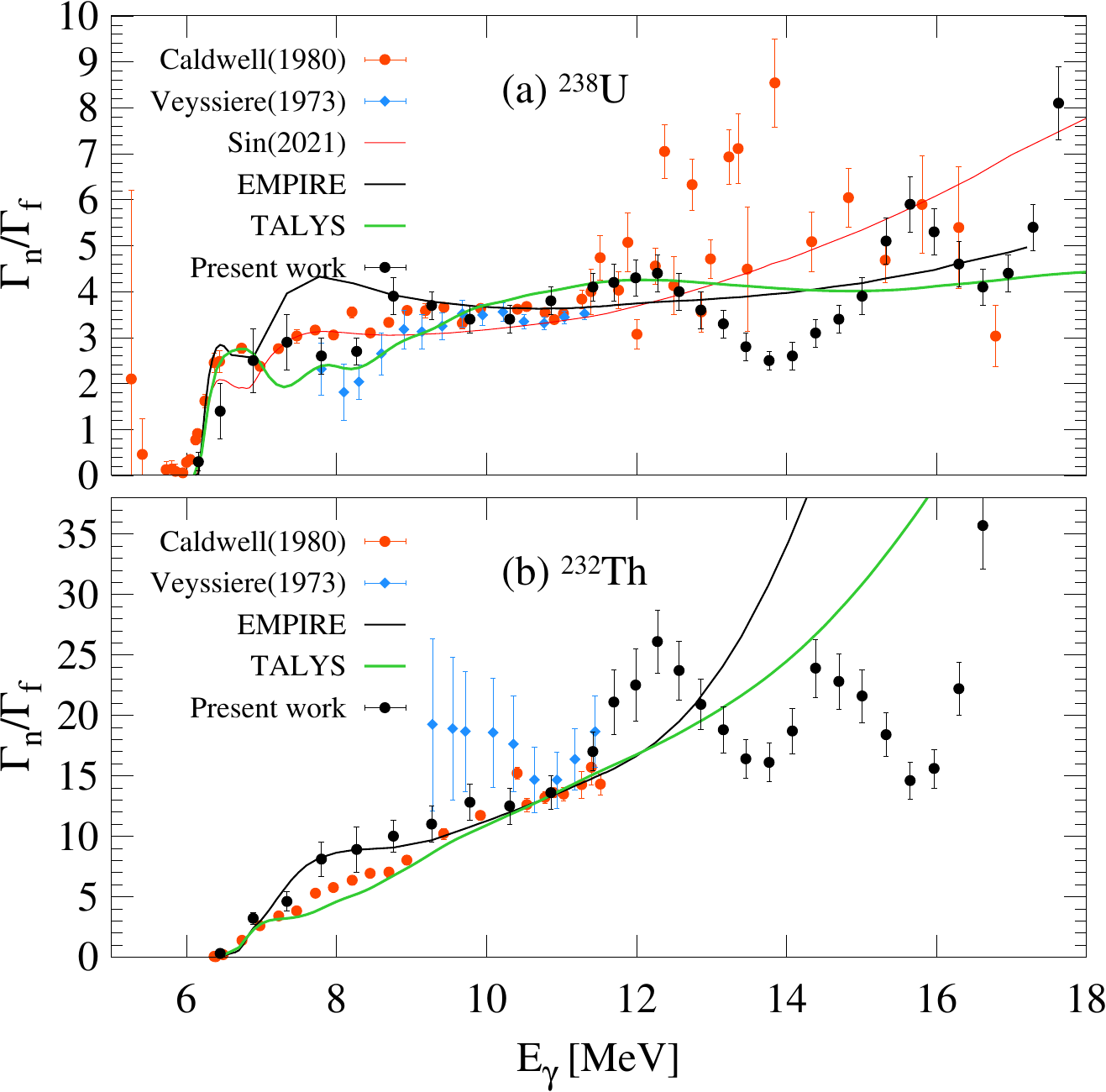} 
\caption{Neutron to fission branching ratio $\Gamma_n/\Gamma_f$ as a function of the excitation energy for (a) $^{238}$U and (b) $^{232}$Th. Present data (full black dots) are compared to the results of Caldwell et al.~\cite{Caldwell1980_PRC} (red empty dots) and of Veyssiere et al.~\cite{Veyssiere1973} (full blue diamonds), and to statistical model calculations: present EMPIRE  (black lines), TALYS (green lines) calculations and the ones of Sin et al.~\cite{msin_2021} (red line). \label{fig_GnGf_fis_chance}}
\end{figure}

The decomposition of the total photofission cross section into the first- and second-chance contributions enables us to extract the neutron to fission branching ratio $\Gamma_n/\Gamma_f$ in an energy range extended above $B_{nf}$. At low excitation energies below $B_{nf}$, $\Gamma_n/\Gamma_f$ can be directly determined from $\sigma_{\gamma,\,n}/\sigma_{\gamma,\,F}$. Above $B_{nf}$, $\Gamma_n/\Gamma_f$ is determined from the ratio of the total photoneutron cross section to the first-chance fission one:
\begin{equation}
\Gamma_n/\Gamma_f = \cfrac{\sigma_{\gamma,\,n}+\sigma_{\gamma,\,2n}+\sigma_{\gamma,\,nf}}{\sigma_{\gamma,\,f}}.
\end{equation}

The results of this analysis are shown in Fig.~\ref{fig_GnGf_fis_chance} for (a) $^{238}$U and (b) $^{232}$Th, where we can see the formation of a plateau up to the energy at which the contribution of the second-chance fission begins to be significant. Its presence suggests that there is a constant proportionality factor between the transition states density in the saddle point and the fundamental deformation level density in the residual nucleus following neutron emission. This observation is in agreement with Bohr's picture~\cite{bohr_1956} according to which the transition states density in the saddle point is similar to the level density of permanently deformed nuclei at fundamental deformation.

\section{Comparison with statistical model calculations}
\label{sec_stat_mod_calc}

The new experimental data are now compared with statistical model calculations obtained by the EMPIRE \cite{herman_2007_empire} and TALYS \cite{koning_2023_talys} codes. Since the entrance channel plays a fundamental role for an accurate description of the various reaction channels, the fit to experimental photoabsorption cross sections has been tested with several Lorentzian-type closed-forms (SLO, MLO1, SMLO) plus the quasi-deuteron contribution for the E1 $\gamma$-ray strength functions~\cite{capote2009_ripl,plujko_2018}. For both $^{238}$U and $^{232}$Th, we found that the SMLO model reproduces best the experimental data, especially in the low-energy region. Table~\ref{table_GDR_par} gives for each nucleus the fit values of the two Lorentzian centroids, widths and peak cross sections and the normalization factor for the quasideuteron photo-absorption cross section used within the EMPIRE code. In contrast, the TALYS code uses the SMLO E1 and M1 photon strength functions in tabulated format \cite{goriely_2019,goriely_2019b}. In particular, with respect to the original tables, correction factors to the overall strength amplitude $f_\sigma$, centroid energy $\Delta_E$ and width $f_\Gamma$ are applied. For $^{238}$U and $^{232}$Th, these correction factors amount to $f_\sigma=0.87$ and 1.22, $\Delta_E=-0.07$ and 0.13~MeV and $f_\Gamma=0.90$ and 0.93, respectively. 

Specific optical model (OM) potentials for $^{238}$U (RIPL ID - 2413) and $^{232}$Th (RIPL ID - 2412)~\cite{capote2009_ripl} have been used in both codes to obtain the transmission coefficients for neutron emission. The level densities, both at the equilibrium deformation and at the saddle points, have been described with the enhanced generalized superfluid model~\cite{herman_2007_empire} in EMPIRE code and with the Hartree-Fock-Bogolyubov (HFB) plus combinatorial model \cite{goriely_2008} in TALYS code. In both approaches, the ground state level densities have been adjusted to reproduce the low-lying discrete level scheme and the $s$-wave resonance spacing at the neutron separation energy. However, since no such constraint exists on the level densities at the saddle points, the corresponding predictions are adjusted directly to optimize the present experimental fission cross sections. 

Within the EMPIRE code, the fission probabilities were computed in the frame of the Optical Model for Fission (OMF)~\cite{bhandari_1979,sin_2016}, which models the coupling between the states at fundamental deformation with the super- and hyper-deformed states as well as the vibrational states damping with the increase in the excitation energy. The fission barriers heights and widths for the first compound nuclei $^{238}$U and $^{232}$Th have been fixed by reproducing the existing fission cross sections in the sub-barrier region, as shown in Fig.~\ref{fig_cs_gF_low}. The properties of the 1$^-$ and 0$^-$ discrete transition states were obtained with EMPIRE by reproducing the sub-barrier fission resonances. 
For both nuclei, we increased in EMPIRE the first neutron emission pre-equilibrium contribution computed within the $\texttt{PCROSS}$ model~\cite{capote_1991_pcross} in order to reproduce the present $(\gamma,\,n)$ cross sections at excitation energies above $\sim$14~MeV. Finally we tuned the fission barrier parameters for the second compound nuclei $^{237}$U and $^{231}$Th in order to reproduce the interplay between $\sigma_{\gamma,\,in}$ with i~=~1~--~3 and $\sigma_{\gamma,\,F}$. The fission barrier parameters used in the present EMPIRE calculations are listed in Table~\ref{table_fis_barr}. We note small differences between the present fission barrier parameters for the U isotopes and the ones obtained by Sin~\emph{et al.} in~Ref.~\cite{msin_2021} by following the Saclay and bremsstrahlung data. 

In contrast, TALYS makes use of the full HFB fission paths obtained with the BSk14 Skyrme interaction \cite{goriely_2007,capote2009_ripl}. In this approach, the probability to penetrate the fission barrier is obtained using the WKB approximation applied to the full HFB fission path, as described in Refs.~\cite{goriely_2009,goriely_2011}. 
To optimize the description of the fission cross sections, the HFB fission paths can be globally renormalized without modifying their topology, simply by scaling the potential energy curve by a given factor $f_B$. The scaling factors obtained for  $^{238}$U and  $^{237}$U amount to $f_B=0.80$ and 0.91, respectively, and for $^{232}$Th and $^{231}$Th to $f_B=0.83$ and 0.98, respectively. The resulting topology of the fission path remains relatively different than the one used in EMPIRE with the parameters given in Table~\ref{table_fis_barr}. Additionally, two free parameters associated with energy shift and a scaling factor of the entropy, are introduced to adjust the HFB plus combinatorial nuclear level densities at both saddle points in both U and Th isotopes.

As shown in Figs.~\ref{fig_cs_partiale}--\ref{fig_csF_fis_chance}, both EMPIRE and TALYS ingredients could be adjusted to reproduce satisfactorily the various cross sections obtained experimentally in the present study despite relatively different descriptions, in particular, of nuclear level densities and fission transmission coefficients. The major weakness probably lies in the prediction of the $^{238}$U first chance fission (Fig.~\ref{fig_csF_fis_chance}) for which the peak cross section around 14~MeV is significantly underestimated, while the second chance fission is overestimated. A similar pattern, though less pronounced, is found for $^{232}$Th.

Concerning the neutron to fission branching ratio (Fig.~\ref{fig_GnGf_fis_chance}), the present results show structures at energies above $B_{nf}$, suggesting a fluctuation around a constant value.  The statistical model calculations show a strong increase in favour of the neutron channel, which may indicate an issue in the level densities used in the statistical model calculations. The difference between the present and the Sin et al.~\cite{msin_2021} $\Gamma_n/\Gamma_f$ results may originate from the fact that, in the present EMPIRE calculations, the pre-equilibrium contribution for the first neutron emission has been increased in order to reproduce the present $\sigma_{\gamma,\,n}$  above the GDR peak, which are higher than the Veyssiere et al.~\cite{Veyssiere1973} ones followed by Sin et al.~\cite{msin_2021}. 

\begin{table}[b]
\caption{\label{table_GDR_par} EMPIRE GDR parameters for $^{232}$Th and $^{238}$U obtained by fitting the present photoabsorption results with a SMLO function described in~Ref.~\cite{plujko_2018}.}
\begin{ruledtabular}
\begin{tabular}{lccccccc}
                         & E(1)           & $\Gamma$(1)      & $\sigma$(1)      & E(2)           & $\Gamma$(2)      & $\sigma$(2)      & QD \\ 
    \textrm{CN}          & \textrm{(MeV)} & \textrm{(MeV)}   & \textrm{(mb)}    & \textrm{(MeV)} & \textrm{(MeV)}   & \textrm{(mb)}    &    \\ \colrule
    \textrm{$^{232}$Th}  & 11.45          & 3.92             & 373.6            & 14.10          & 3.75             & 359.9            & 1. \\
    \textrm{$^{238}$U}   & 10.91          & 2.93             & 308.1            & 14.03          & 4.50             & 359.3            & 3. \\
\end{tabular}
\end{ruledtabular}
\end{table}

\begin{table*}
\caption{\label{table_fis_barr} Fission barrier parameters used in present EMPIRE calculations. $V_1(\hbar \omega_1)$, $V_2(\hbar \omega_2)$ and $V_3(\hbar \omega_3)$ are the fission barrier heights~(curvatures). $V_\mathrm{II}(\hbar \omega_\mathrm{II})$ and $V_\mathrm{III}(\hbar \omega_\mathrm{III})$ are second and third well heights (curvatures) at super- and hyper-deformations \footnote{all values are given in MeV.}.}
\begin{ruledtabular}
\begin{tabular}{ccccccccccccc}
 Nucleus      & $V_1$ & $\hbar \omega_1$ & $V_\mathrm{II}$ & $\hbar \omega_\mathrm{II}$ & $V_2$ & $\hbar \omega_2$ & $V_\mathrm{III}$ & $\hbar \omega_\mathrm{III}$ & $V_3$ & $\hbar \omega_3$ \\ \hline
 $^{238}$U    & 6.25  & 1.00             & 1.30            & 1.00                       & 5.70  & 0.60             &                  &                             &       &                  \\     
 $^{237}$U    & 5.35  & 0.50             & 2.30            & 1.00                       & 5.85  & 1.50             & 5.57             & 1.00                        & 5.85  & 1.50             \\  
 $^{232}$Th   & 5.80  & 0.70             & 4.75            & 0.30                       & 6.00  & 1.70             & 5.20             & 0.90                        & 6.00  & 1.70             \\    
 $^{231}$Th   & 6.05  & 0.70             & 3.00            & 1.00                       & 6.20  & 1.50             &                  &                             &       &                  \\ 
\end{tabular}
\end{ruledtabular}
\end{table*}
 
\section{Summary and outlook}
\label{sec_summary}

New measurements of photoneutron and photofission reactions on $^{238}$U and $^{232}$Th have been performed in the GDR energy region using 5.87~MeV~--~20.14~MeV quasi-monochromatic laser Compton scattering $\gamma$-ray beams of the NewSUBARU synchrotron radiation facility. A high-and-flat efficiency (FED) moderated $^3$He detection array together with an associated neutron-multiplicity sorting method detailed in Sec.~\ref{sec_NMS} have been employed. 

We obtained the cross sections for the $(\gamma,\,n)$, $(\gamma,\,2n)$, $(\gamma,\,3n)$ photoneutron reactions and for the photofission reaction, as well as prompt-fission-neutron quantities, i.e. mean numbers of PFNs, width of PFNs multiplicity distribution and average energies of PFNs. The experimental data have been interpreted under reasonable assumptions in order to extract first- and second-chance fission contributions. This made possible the determination of the neutron to fission branching ratio $\Gamma_n/\Gamma_f$~(Fig.~\ref{fig_GnGf_fis_chance}). We show that the present PFN data are well reproduced by Los Alamos calculations. The photoneutron and the photofission experimental results have been well reproduced with both EMPIRE and TALYS calculations by slight adjustments of model parameters. All the experimental results obtained in the present paper are available in numerical format in the Supplemental Material~\cite{supplemental_material}.

The present LCS $\gamma$-ray beam data obtained with a neutron detector of Livermore type together with the Saclay and Livermore positron in flight annihilation data represent the only monochromatic $\gamma$-ray beam measurements on $^{238}$U and $^{232}$Th in the GDR energy range. We obtained a good agreement with the Saclay results for $^{238}$U and with the Livermore ones for $^{232}$Th. This is contradictory to the idea of finding a systematic solution to the discrepancies between the Saclay and Livermore data and points to the need to remeasure the photoneutron cross sections over a wide range of nuclei and incident energies using quasi-monochromatic $\gamma$-ray beams such as the ones currently available at the SLEGS~\cite{wang_fan_2022} and HI$\gamma$S~\cite{litvinenko_1997} LCS facilities.

An interpretation of the present experimental measurements based on a Bayesian statistics approach with no assumptions on the shape of the PFNs multiplicity distributions is currently ongoing~\cite{stopani_unpublished}. The analysis aims to provide independent cross sections of emission of 1, 2, 3, $\dots$ prompt neutrons in photofission reaction.

\section{Acknowledgments}
The authors are grateful to H. Ohgaki of the Institute of Advanced Energy, Kyoto University for making a large volume LaBr$_3$(Ce) detector available for the experiment and sincerely thank M.~Bj{\o}r{\o}en of the University of Oslo for helping with the experiment. 
The authors express their gratitude to Prof. M.~Sin of the University of Bucharest, Romania, for making available her statistical model calculation results in numerical format, for her interest in this paper and valuable discussions. 
D.F., I.G. and A.T. acknowledge the support from the Romanian Ministry of Research, Innovation and Digitalization/Institute of Atomic Physics from the National Research - Development and Innovation Plan III for 2015--2020/Programme 5/Subprogramme 5.1 ELI-RO, project GANT-Photofiss No 14/16.10.2020. 
D.F and I.G acknowledge the support from the Romanian Project PN-23-21-01-02. 
S.G. acknowledges financial support from the Fonds de la Recherche Scientifique (F.R.S.-FNRS) and the Fonds Wetenschappelijk Onderzoek - Vlaanderen (FWO) under the EOS Projects nr O022818F and O000422F. 

\appendix

\section{Background reaction rates from Al container and O}\label{annex_B}

\begin{figure}[t]
\includegraphics[width=0.49\textwidth, angle=0]{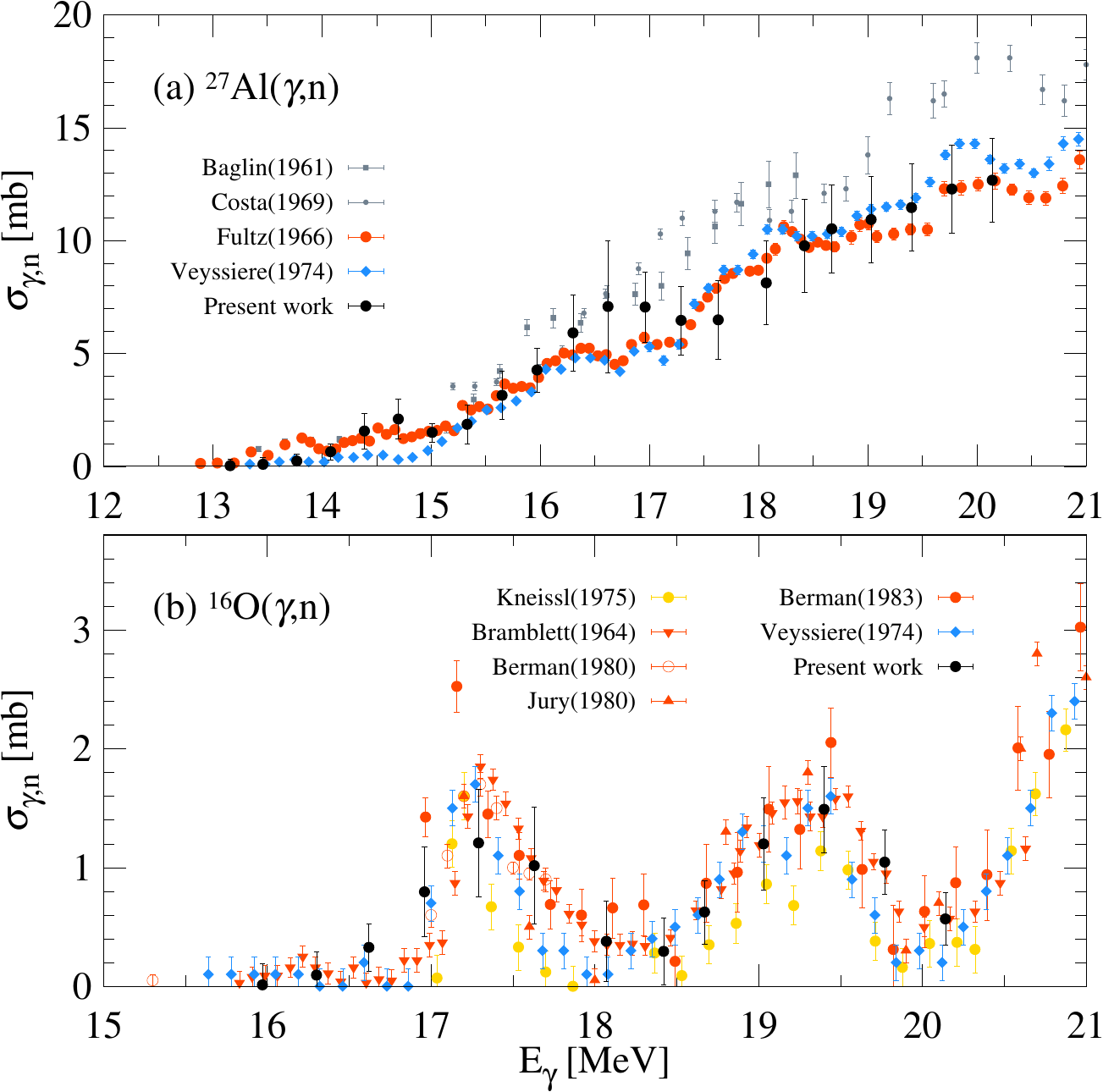} 
\caption{Present photoneutron cross sections (full black dots) for (a) $^{27}$Al and (b) $^{16}$O compared with positron in flight annihilation data taken at Saclay~\cite{Veyssiere1974} (full blue diamonds), at Livermore~\cite{Fultz1966,Bramblett1964,Berman1980,Berman1983,Jury1980} (red symbols) and at Giessen~\cite{Kneissl1975} (full yellow dots), and with bremsstrahlung data~\cite{Baglin1961,Costa1969} (gray symbols).  \label{fig_cs_Al_O16}}
\end{figure}

We observed non-zero net counts of single neutron events in the empty Al container and water target measurements. The present $^{27}$Al$(\gamma,\,n)$ and $^{16}$O$(\gamma,\,n)$ cross sections computed based on the these background measurements are shown in Fig.~\ref{fig_cs_Al_O16}. There is reasonable agreement between present and the preceding results obtained with bremsstrahlung beams~\cite{Baglin1961,Costa1969} and with positron in flight annihilation photon beams at Saclay~\cite{Veyssiere1974}, Livermore~\cite{Fultz1966,Bramblett1964,Berman1980,Berman1983,Jury1980} and Giessen~\cite{Kneissl1975}.  

In order to subtract the corresponding $\Delta n$ contributions to the actinide targets measurements, we sum the background contributions following a normalization for the number of incident photons and surface concentration of Aluminum and, respectively Oxygen nuclei:
\begin{equation}
\Delta n = n_\texttt{w} \cdot \cfrac{N_{\gamma \texttt{t}}}{N_{\gamma \texttt{w}}} \cfrac{N_{O \texttt{t}}}{N_{O \texttt{w}}} + n_\texttt{a} \cdot \cfrac{N_{\gamma \texttt{t}}}{N_{\gamma \texttt{a}}}. 
\end{equation}
Here, $n_\texttt{a}$ and $n_\texttt{w}$ are the net numbers of single-neutron events recorded for the Al and water target measurements, respectively. $N_{\gamma \texttt{t}}$, $N_{\gamma \texttt{a}}$ and $N_{\gamma \texttt{w}}$ are the numbers of incident photons for the actinide target, empty Al container and respectively water target measurements. $N_{O \texttt{t}}$ and $N_{O \texttt{w}}$ are the surface concentrations of Oxygen nuclei in the actinide and respectively water targets. Their ratio $N_{O \texttt{t}}/N_{O \texttt{w}}$ is equal to  0.139 for the U$_3$O$_8$ target and 0.235 for the ThO$_2$ target. A normalization for the number of Al nuclei was not necessary, as identical Al containers were used for the actinide target encapsulation and for the empty container measurements. The $\Delta n$ estimated sum background contribution from Aluminum and Oxygen nuclei is computed separately for each actinide target, for the inner ring and summed outer rings and for each experimental point at incident energies above the $^{27}$Al and $^{16}$O neutron separation energies, at 13.1~MeV and respectively 15.7~MeV. 

\section{Statistical treatment of multiple firing photoneutron and photofission reactions}\label{annex_A}

The statistical treatment of multiple firing of competitive $(\gamma,\,in)$ photoneutron reactions originally described in Ref.~\cite{Gheorghe2017} is here updated to account also for photofission reactions. 

The probability for an incident LCS $\gamma$-ray pulse to induce any combination of $r$~$\ge$~1 photoneutron and/or photofission reactions is defined as:
\begin{equation}
\mathbf{P}(r)=\sum_{k=1}^{k_{max}} \mathcal{P}(\mu_k,r) \cdot p_w(k),
\end{equation}
where $k$ is the time interval index introduced in Sec.~1.1 of Ref.~\cite{Gheorghe2021} and $p_w(k)$ is the corresponding weight factor given by Eq.~(3) of Ref.~\cite{Gheorghe2021}. As described in Refs.~\cite{Gheorghe2021,Utsunomiya18}, in order to account for the electron beam current exponential decay with time, the total irradiation is divided into $k_{max}$ shorter time intervals, each with an approximately constant $\langle N_{\gamma}(k) \rangle$ mean number of Poisson distributed photons per pulse. Thus, for a given time interval $k$, the number of reactions induced by a $\gamma$-ray pulse also follows a Poisson distribution $\mathcal{P}(\mu_k,r)$ with mean value given by:
\begin{equation}
\mu_k=\mathbf{P}\cdot\langle N_{\gamma}(k) \rangle,
\end{equation}
where $\mathbf{P}$ is the probability for each photon to induce a photoneutron or a photofission reaction:
\begin{equation}
\mathbf{P} =  \sum_{x=1}^\texttt{N} ( \sigma_{\gamma,\,xn} + \sigma_{\gamma,\,fxn} )\cdot n_T\xi.
\end{equation}
For simplicity of notation, we have cycled $\sigma_{\gamma,\,xn}$ up to the highest significant PFN emission multiplicity $\texttt{N}$ with zero values for $x$ higher than the maximum photoneutron multiplicity energetically allowed $\texttt{x}$. The $\sigma_{\gamma,\,fxn}$ cross section for photofission with emission of $x$ PFNs is related in Eq.~\eqref{eq_cs_gfin_cs_gF} to the total $\sigma_{\gamma,\,F}$ photofission cross section.

Let us consider the 
\begin{equation}\label{eq_ref_rx_fx}
(r_xf_x)=(r_1,r_2,\dots,r_\texttt{N},f_1,f_2,\dots,f_\texttt{N})
\end{equation}
combination of reactions which sum up to
\begin{equation}\label{eq_ref_def_r_total}
r =\sum_{x=1}^\texttt{N} (r_x+f_x),
\end{equation}
where $r_x$ is the number of $(\gamma, \,xn)$ reactions and $f_x$ the number of $(\gamma, \,fxn)$ reactions. Given that $r$ reactions were induced by an incident $\gamma$-ray pulse, the conditional probability for firing the given $(r_xf_x)$ combination is:
\begin{equation}
\mathcal{R}\Big((r_xf_x) \big|r =\sum_{x=1}^\texttt{N} (r_x+f_x)\Big) = r! \prod_{x=1}^\texttt{N} \cfrac{P_x^{r_x}F_x^{f_x}}{r_x!f_x!},  
\end{equation}
where $P_x$ is the $(\gamma, \,xn)$ branching ratio given by:
\begin{equation}
P_x={ \sigma_{\gamma,\,xn}} / {\sum_{x=1}^{\texttt{N}} (\sigma_{\gamma,\,xn}+\sigma_{\gamma,\,fxn})}
\end{equation}
and $F_x$ is the  $(\gamma, \,fxn)$ branching ratio:
\begin{equation}
F_x={ \sigma_{\gamma,\,fxn}} / {\sum_{x=1}^{\texttt{N}} (\sigma_{\gamma,\,xn}+\sigma_{\gamma,\,fxn})}.
\end{equation}

Then, the unconditional probability for a $\gamma$-ray pulse to induce the sequence of $(r_xf_x)$ reactions is:
\begin{equation}
\mathbf{R}\Big((r_xf_x)\Big)= \mathcal{R}\Big((r_xf_x) \big|r =\sum_{x=1}^\texttt{N} (r_x+f_x)\Big) \cdot \mathbf{P}(r)
\end{equation}
Assuming that $(r_xf_x)$ reactions have been fired, the conditional probability to detect the $(d_{rx}d_{fx})$ combination of $d_{rx}$ neutrons emitted in $(\gamma, \,xn)$ reactions and $d_{fx}$ neutrons emitted in $(\gamma, \,fxn)$ reactions where $x$ takes values from 1 to $\texttt{N}$, is given by:
\begin{align}
\mathcal{D}\Big((d_{rx}d_{fx}) \big| (r_xf_x)\Big) = & \prod_{x=1}^\texttt{N} {}_{x\cdot r_x}C_{d_{rx}} \, \varepsilon_{\gamma,xn}^{d_{rx}}(1-\varepsilon_{\gamma,xn})^{x\cdot r_x - d_{rx}} \cdot \nonumber \\
 & \cdot  {}_{x\cdot f_x}C_{d_{fx}} \, \varepsilon_{\gamma,f}^{d_{fx}}(1-\varepsilon_{\gamma,f})^{x\cdot f_x - d_{fx}}. 
\end{align}
$\varepsilon_{\gamma,xn}=\varepsilon(E_{\gamma,\,xn})$ is the efficiency of detecting $(\gamma, \,xn)$ photoneutrons of $E_{\gamma,\,xn}$ average energy. $\varepsilon_{\gamma,f}=\varepsilon(E_{\gamma,\,f})$ is the efficiency of detecting neutrons emitted in the $(\gamma, \,fxn)$ photofission reactions, where we have assumed a constant average energy of PFNs, regardless of the $x$ PFN emission multiplicity.

Now we can define the unconditional probability for one $\gamma$-ray pulse to induce the $(r_xf_x)$ sequence of reactions and the $(d_{rx}d_{fx})$ combination of neutrons to be recorded:
\begin{equation}
\mathbf{D}\Big((r_xf_x);(d_{rx}d_{fx})\Big) = \mathbf{R}\Big((r_xf_x)\Big) \cdot \mathcal{D}\Big((d_{rx}d_{fx}) \big|(r_xf_x)\Big).
\end{equation}
However, $\mathbf{D}\Big((r_xf_x);(d_{rx}d_{fx})\Big)$ must be linked to the experimentally observed $\mathbf{D}_i$ probability of recording $i$-fold neutron events, regardless of the precise $(r_xf_x)$ firing and $(d_{rx}d_{fx})$ detection configurations:
\begin{equation}
\mathbf{D}_i = \sum_{\sum_{x=1}^{\texttt{N}}(d_{rx}+d_{fx})=i} \mathbf{D}\Big((r_xf_x);(d_{rx}d_{fx})\Big).
\end{equation}
Finally, we obtain the {\it multiple firing} scenario calculated values for the:
\begin{itemize}
\item $N^{MF}_i$ $i$-fold cross sections, obtained as: 
\begin{equation}
N^{MF}_i=\mathbf{D}_i/{n_T \xi}; 
\end{equation}
\item $E^{MF}_i$ average energy of neutrons recorded in $i$-fold events, expressed as:
\begin{equation}
E^{MF}_i = \cfrac{ \sum_{\sum(d_{rx}+d_{fx})=i} \mathbf{D}\Big((r_xf_x);(d_{rx}d_{fx})\Big) \mathrm{E}\,\Big((d_{rx}d_{fx})\Big)}{\mathbf{D}_i}, 
\end{equation}
where 
\begin{equation}
\mathrm{E}\,\Big((d_{rx}d_{fx})\Big)=\cfrac{\sum_{x=1}^\texttt{N} d_{rx} E_{\gamma,\,xn}+d_{fx}E_{\gamma,\,f} }{\sum_{x=1}^\texttt{N} d_{rx}+d_{fx}}
\end{equation}
is the average energy for the $(d_{rx}d_{fx})$ combination of detected neutrons.
\end{itemize}

\bibliographystyle{apsrev4-1}
\bibliography{newsubarubibfile}

\end{document}